\documentclass[a4paper]{article}

\usepackage[margin=1in]{geometry} 

\usepackage{indentfirst}
\usepackage{caption}
\usepackage{amsmath}
\usepackage{amsthm}
\usepackage{amssymb}
\usepackage{titling}
\usepackage[utf8]{inputenc}

\usepackage[sort&compress,numbers,square]{natbib}
\theoremstyle{plain}

\theoremstyle{definition}

\usepackage{letltxmacro}
\LetLtxMacro{\originaleqref}{\eqref}
\renewcommand{\eqref}{Eq.~\originaleqref}

\usepackage{graphicx, color}
\graphicspath{{fig/}}

\usepackage{algorithm, algpseudocode} 
\usepackage{mathrsfs} 

\usepackage{lipsum}

\title{Modified Kirchhoff's Laws for Electric-Double-Layer Charging in Porous Media}
\author{Filipe Henrique$^1$ \and Pawel J. Zuk$^{2,3}$ \and Ankur Gupta$^1$\thanks{corresponding author. E-mail: ankur.gupta@colorado.edu.}}

\date{\small
	$^1$Department of Chemical and Biological Engineering, University of Colorado, Boulder\\
	$^2$Institute of Physical Chemistry, Polish Academy of Sciences, Kasprzaka 44/52, 01-224 Warsaw, Poland\\
    $^3$ Department of Physics, Lancaster University, Lancaster LA1 4YB, United Kingdom
}

\begin{document}
	\maketitle
	
	\begin{abstract}
		Understanding the dynamics of electric-double-layer (EDL) charging in porous media is essential for advancements in next-generation energy storage devices. Due to the high computational demands of direct numerical simulations and a lack of interfacial boundary conditions for reduced-order models, the current understanding of EDL charging is limited to simple geometries. Here, we present a theoretical framework to predict EDL charging in arbitrary networks of long pores in the Debye-Hückel limit without restrictions on EDL thickness and pore radii. We demonstrate that electrolyte transport is described by Kirchhoff's laws in terms of the electrochemical potential of charge (the valence-weighted average of the ion electrochemical potentials) instead of the electric potential.  By employing this equivalent circuit representation with modified Kirchhoff's laws, our methodology accurately captures the spatial and temporal dependencies of charge density and electric potential, matching results obtained from computationally intensive direct numerical simulations. Our framework provides results up to five orders of magnitude faster, enabling the efficient simulation of thousands of pores within a day. We employ the framework to study the impact of pore connectivity and polydispersity on electrode charging dynamics for pore networks and discuss how these factors affect the timescale, energy density, and power density of the capacitive charging. The scalability and versatility of our methodology make it a rational tool for designing 3D-printed electrodes and for interpreting geometric effects on electrode impedance spectroscopy measurements.\\
		
	\end{abstract}

\section{Introduction}

Ionic transport in dilute electrolytes has been successfully predicted through the Poisson-Nernst-Planck (PNP) equations and their derivatives \cite{bazant2004diffuse,ho2005electrolytic,muthukumar2006simulation,davidson2014chaotic,jubin2018dramatic}. Given the central role that ionic transport in porous media plays in energy storage devices such as electrochemical capacitors and batteries \cite{simon2008materials,simon2020perspectives}, significant effort has been devoted to developing continuum models that describe this phenomenon \cite{sakaguchi2007charging,biesheuvel2010nonlinear,mirzadeh2014enhanced,schmuck2015homogenization}. In these systems, electric double layers store energy through the electrostatic attraction of counterions to the electrode surfaces. Since the capacitance of electric double layers increases with surface area, the electrodes are usually porous and consist of surface areas as high as 3300 m$^2$/g \cite{jiang2021large}.
One of the exciting developments in the area is the 3-D printing of supercapacitor electrodes, with applications to the Internet of Things \cite{simon2020perspectives} and to wearable energy storage \cite{chu20213d}. However, physical principles for the rational design of pore network geometries with optimal performance remain an open question.

Recent studies have demonstrated that molecular dynamics simulations provide full descriptions of electric-double-layer charging in confinement, including non-idealities due to concentrated electrolytes, high potentials, and surface chemistry \cite{kondrat2023theory}. However, despite these capabilities, their high computational cost precludes the study of large systems with complex pore networks and realistically long timescales \cite{lahrar2021carbon,wu2022understanding}. Though more computationally favorable than molecular dynamics, modified PNP equations \cite{kilic2007steric1,kilic2007steric2,storey2012effects,xu2014self,gupta2020ionic,de2020interfacial} and dynamic density functional theory \cite{kondrat2014accelerating,tomlin2022impedance} are also challenging to employ in porous media due to complexities in the geometry.

\par{} In view of these shortcomings, transmission-line (TL) models stemming from the seminal work of de Levie \cite{de1963porous,de1964porous} continue to be used as a good qualitative guide of the physics of electrolyte charging in experimental investigations \cite{black2010pore,movskon2021transmission}. Their use can be justified by similar diffusion-like equations found in more detailed dynamic density functional theory models \cite{kondrat2023theory} and good matches obtained with experimental results using fitting parameters \cite{gebbie2013ionic,wu2022understanding,bi2020molecular, zeng2021modeling}.
\par{}Nevertheless, the majority of the TL models are conceived as single-pore models. de Levie's model, for instance, may be formally justified by a linearization of the PNP equations for a single pore at low electric potentials \cite{wu2022understanding} in the thin-double-layer limit \cite{janssen2021transmission}. Variations of the TL model account for high potentials \cite{biesheuvel2010nonlinear}, surface conduction \cite{mirzadeh2014enhanced}, arbitrary double-layer thickness \cite{gupta2020charging,henrique2022charging}, ionic diffusivity asymmetry \cite{henrique2022impact}, or a stack-electrode model \cite{lian2020blessing}. However, none of these variations capture the geometric effect of pore connections or pore sizes beyond a lumped-parameter approach. 

\par{}In our view, this crucial knowledge gap exists because of two primary reasons. First, the direct numerical simulation of PNP in an arbitrary network of pores is computationally infeasible. Second, while it is understood that the dynamics of double-layer charging in single pores differs between the thin- and the overlapping-double-layer limits \cite{gupta2020charging,henrique2022charging}, the interaction of pores of different sizes remains undescribed.  
\par{} In this article, we devise a comprehensive theoretical framework to predict electric-double-layer charging of a binary electrolyte in arbitrary networks of long pores in the  Debye-Hückel limit. The proposed approach provides reduced-order transport equations in each pore and describes voltage and charge relationships across junctions and loops of pores, i.e., effective Kirchhoff's laws for electrolyte transport in porous media that close the system of equations. We emphasize that these modified Kirchhoff's laws are required to capture the simultaneous effects of diffusion and electromigration, whereas the original Kirchhoff's laws are valid only for purely electromigrative transport. These effective Kirchhoff's laws are written for the electrochemical potential of charge, i.e., for the valence-weighted average of the ion electrochemical potentials, instead of the electric potential. We compare our approach against direct numerical simulations of the PNP equations for a range of different geometries and demonstrate that our approach is able to recover the spatial and temporal dependencies of charge density and electric potential. The TL model devised is computationally inexpensive and enables the simulation of a network consisting of thousands of pores in less than a day whereas our direct numerical simulations of the full PNP equations take upwards of a month for a 3-pore network using 28 cores (see the Methods section). The scalability of the proposed TL methodology enables us to quantify the impact of pore size distribution and pore connectivity, and consequently to uncover guiding principles for optimizing the design of porous electrodes.

\begin{figure*}[b!]
    \centering
    \includegraphics[width=\textwidth]{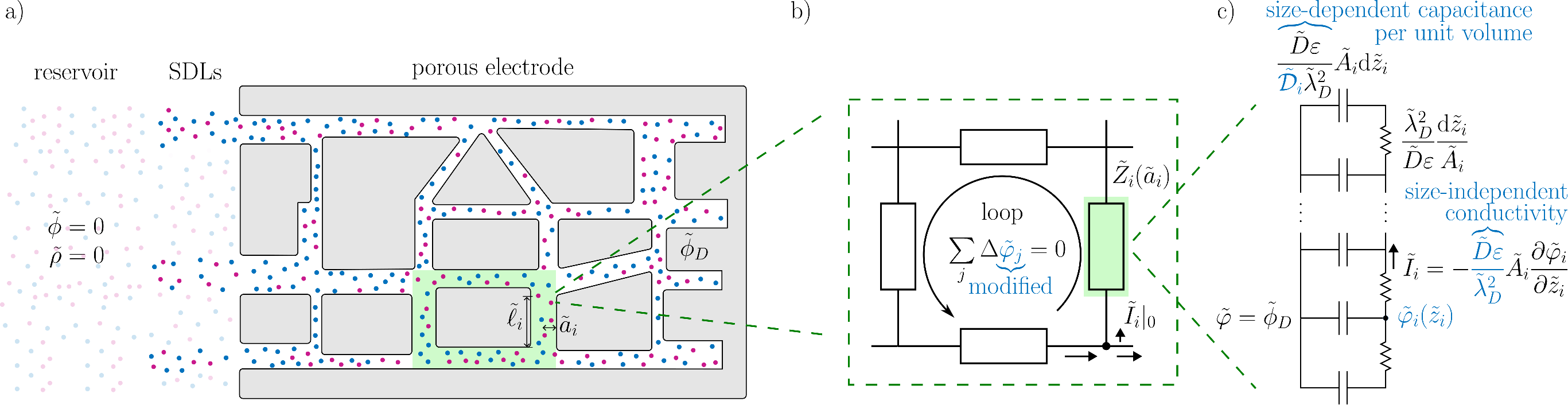}
    \caption{\textbf{Transmission-line (TL) model for electric-double-layer charging in networks of long pores}. a) Schematic of an arbitrary porous electrode geometry representable by a network of long pores, with cations in purple and anions in blue. The electroneutral reservoir (transparent ions) has a reference potential $\tilde{\phi}=0$, connected through static diffusion layers (opaque ions outside the electrode) to the inlet pores. The perfectly conducting electrode is at a  potential $\tilde{\phi}_D$. The pores are assumed to be cylindrical. Slit pores are addressed in the Supporting Information. The length, radius, and area of the $i$-th pore are respectively $\tilde{\ell}_i$, $\tilde{a}_i$, and $\tilde{A}_i$.  b) A group of pores represented by a circuit loop. The balance of diffusion and electromigration requires effective forms of Kirchhoff's laws written in terms of the electrochemical potential of charge $\tilde{\varphi}_i=\tilde{\phi}_i + \tilde{\lambda}_D^2\tilde{\rho}_i/\varepsilon$ of each pore, which is continuous across junctions as seen from  \eqref{eq:kvl}. The impedance of each pore is written as $\tilde{Z}_i(\tilde{a}_i)$ and the current at $\tilde{z}_i=0$ by $\tilde{I}_i|_0$. c) Representation of each pore of arbitrary double-layer thickness in the transmission line model with a pore-size dependent impedance consistent with Ref. \cite{henrique2022charging} for isolated pores. Each resistor has the same conductivity $\tilde{D}\varepsilon/\tilde{\lambda}_D^2$ and each capacitor has a capacitance per unit volume $\tilde{D}\varepsilon/(\mathcal{D}_i\tilde{\lambda}_D^2)$.}
    \label{fig:schematic}
\end{figure*}
  
\section{Results and Discussion}
\subsection{Charging Dynamics}
Throughout the article, we represent dimensional variables by tildes to distinguish them from their dimensionless counterparts. We consider a network of $N$ long cylindrical pores within a perfectly conducting and ideally blocking electrode at a potential $\tilde{\phi}_D$ relative to an ion reservoir. The reservoir is connected to the inlet pores through static diffusion layers (SDLs), as illustrated in Fig. \ref{fig:schematic}a. The pore network is arbitrary, allowing us to study the effects of pore size distribution, pore connectivity, and spatial arrangement. We assume that the porous network is filled with a binary and symmetric electrolyte with ion diffusivity $\tilde{D}$, though it is straightforward to extend our work to asymmetric ion diffusivities and valences \cite{henrique2022impact}. 

Variables indexed by $i$ pertain to the $i$-th pore, with dimensionless radial and axial coordinates $\tilde{r}_i\in [0,\tilde{a}_i]$ and $\tilde{z}_i\in [0,\tilde{\ell}_i]$. $\tilde{a}_i/\tilde{\ell}\ll 1$ is the long-pore requirement, where $\tilde{\ell}$ is a characteristic length of all pores. The reservoir concentration $\tilde{c}_\infty$ of either ionic species is taken as the concentration scale, and thus the Debye length takes the form $\tilde{\lambda}_D=\sqrt{\varepsilon kT/(2e^2\tilde{c}_\infty)}$, where $kT/e$ is the thermal voltage and $\varepsilon$ is the electrolyte permittivity. We define the relative pore size parameter $\kappa_i=\tilde{a}_i/\tilde{\lambda}_D$. The key question that we seek to address is how pores of different sizes interact to collectively encode an effective charging timescale in the network. To answer this question, we derive a TL model from the PNP equations.

The charge fluxes in the Debye-Hückel regime are produced by gradients of the electrochemical potentials of the ions (see Supplementary Information), 
\begin{equation}
    \tilde{\mathbf{J}}=-\dfrac{\tilde{D}\varepsilon}{\tilde{\lambda}_D^2}\tilde{\nabla}\left(\dfrac{\tilde{\mu}_+-\tilde{\mu}_-}{2e}\right),
\end{equation}
where $\tilde{\mu}_+$ and  $\tilde{\mu}_-$ are the electrochemical potentials of cation and anion, respectively. Due to its role as the effective potential for charge transport, we define electrochemical potential of charge $\tilde{\varphi}=(\tilde{\mu}_+-\tilde{\mu}_-)/(2e)$, where the dimensions of $\tilde{\varphi}$ are purposefully kept the same as $\tilde{\phi}$ to make direct comparisons between the two when deriving effective Kirchhoff's law. In the low-applied-potential limit, $\tilde{\varphi}$ takes the asymptotic form (see Supplementary Information)
\begin{equation}
    \tilde{\varphi}=\tilde{\phi}+\dfrac{\tilde{\lambda}_D^2}{\varepsilon}\tilde{\rho},
\end{equation}
where the prefactor $\tilde{\lambda}_D^2/\varepsilon$ can be interpreted as the required unit conversion from charge density to electric potential, as seen from Poisson's equation.
The long-pore condition allows us to invoke radial equilibrium (i.e., no radial flux) at each time to impose the constancy of $\tilde{\varphi}$, thus constraining the radial dependencies of the charge density $\tilde{\rho}_i$ and the potential $\tilde{\phi}_i$ by the relation
\begin{equation}
\tilde{\varphi}_i(\tilde{z}_i,\tilde{t})=\tilde{\phi}_i(\tilde{r}_i,\tilde{z}_i,\tilde{t})+\dfrac{\tilde{\lambda}_D^2}{\varepsilon}\tilde{\rho}_i(\tilde{r}_i,\tilde{z}_i,\tilde{t})=\bar{\tilde{\phi}}_i(\tilde{z}_i,\tilde{t})+\dfrac{\tilde{\lambda}_D^2}{\varepsilon}\bar{\tilde{\rho}}_i(\tilde{z}_i,\tilde{t}),
\label{eq:rho_phi_eq}
\end{equation}
where bars represent cross-sectional averages. \eqref{eq:rho_phi_eq} could also be understood as the Boltzmann distribution in the radial direction, linearized for low potentials. The axial charge flux is proportional to $ - \frac{\partial}{\partial \tilde{z}_i} \left( \tilde{\phi}_i + \frac{\tilde{\lambda}_D^2}{\varepsilon}\tilde{\rho}_i \right) $, where the term with -$\frac{\partial \tilde{\phi}_i}{\partial \tilde{z}_i}$ represents the electromigrative flux and the term with -$\frac{\partial \tilde{\rho}_i}{\partial \tilde{z}_i}$ describes the diffusive flux. For the thin-double-layer limit, i.e., $\kappa_i \gg 1$, the average charge density is such that $\bar{\tilde{\rho}}_i/(\varepsilon/\tilde{\lambda}_D^2)\sim \tilde{\phi}_D/\kappa_i \ll \tilde{\phi}_D$ in the charged regions and the diffusive charge flux can be neglected. The thin-double-layer limit thus greatly simplifies the analysis and has been widely explored in the literature \cite{de1963porous, de1964porous, mirzadeh2014enhanced, biesheuvel2010nonlinear, biesheuvel2011diffuse}. For an arbitrary $\kappa_i$, however, one cannot further simplify the axial flux as both diffusion and electromigration could be important. 

The relative importance of electromigration and diffusion could change across a junction, depending on the radii of the connected pores. As an example, let us consider a simple scenario with only two pores connected by a junction: one pore has a radius in the thin-double-layer limit such that $\kappa_i \gg 1$, and the other pore has a radius in the overlapping-double-layer limit such that $\kappa_j \ll 1$. For the pore with a thin double layer, as the charge density and the potential screening are restricted to the double layer, $\bar{\tilde{\phi}}_i\sim \tilde{\phi}_D/\kappa_i\ll \tilde{\phi}_D$ and $\bar{\tilde{\rho}}_i\tilde{\lambda}_D^2/\varepsilon\sim \tilde{\phi}_D/\kappa_i\ll  \tilde{\phi}_D$. However, for the one with an overlapping double layer, as charges are present throughout the cross-section, $\bar{\tilde{\rho}}_j\tilde{\lambda}_D^2/\varepsilon \sim \tilde{\phi}_D$ and $\bar{\tilde{\phi}}_j\sim \tilde{\phi}_D$. As evident from the above discussion, the steady-state charge and potential distributions in the pores will be different. Therefore, the junction will present sharp changes in potential and charges to adhere to the individual behavior of the pores \cite{gupta2020charging, henrique2022charging, henrique2022impact}. The preceding discussion assumes only two pores, but in reality, there could be a large number of connected pores and multiple junctions, which could further complicate the analysis.

The complexity of the charge and electric potential profiles across junctions is overcome by the usage of the electrochemical potential of charge to represent charge transport. $\tilde{\varphi}_i$ does not vary along a cross-section -- see \eqref{eq:rho_phi_eq} -- or across a junction (see Supporting Information for a derivation). The continuity of the electrochemical potential of charge across junctions is similar to the continuity of electric potential across a node of an electric circuit  \cite{alexander2013fundamentals}. It is also similar to the continuity of temperature across an interface \cite{deen1998analysis}. That is, while $\bar{\tilde{\phi}}_i$ and $\bar{\tilde{\rho}}_i$ could present sharp changes across a junction, $\tilde{\varphi}_i$ is continuous throughout the region. As such, $\tilde{\varphi}_i$ is a natural quantity to describe the transport of a symmetric electrolyte inside a porous network. To the best of our knowledge, this result, which forms the basis of our methodology, has not been reported before.

The usage of $\tilde{\varphi}_i$ simplifies the description of double-layer charging in a porous network in two primary ways. First, the axial flux is given by $-\frac{\tilde{D}\varepsilon}{\tilde{\lambda}_D^2}\frac{\partial \tilde{\varphi}_i}{\partial \tilde{z}_i}$ and thus the dimensionless conductivity of all the pores is equal. We underscore that this wouldn't be true if we had defined conductivity based on electric potential, as the conductivity would need to be adjusted depending on $\kappa_i$. Second, $\tilde{\varphi}_i$ is continuous across a junction and can be used in TL circuit representations without introducing circuit elements representing the junctions. On the other hand, $\bar{\tilde{\phi}}_i$ would be modeled as discontinuous since it can change across junctions due to different extents of double-layer screening as a function of $\kappa_i$. Though ionic electrochemical potentials have been used as the thermodynamic force driving electroosmotic flow in confinement \cite{peters2016analysis,alizadeh2017multiscale}, the importance of the effective electrochemical potential of charge to TL representations of double-layer charging in porous media had not been recognized.

Before presenting the junction boundary conditions, we briefly describe the transport equations for each pore, which reduce to (see Supporting Information for details)
\begin{equation}
\dfrac{\partial \tilde{\varphi}_i}{\partial \tilde{t}}=\tilde{\mathcal{D}}_i\dfrac{\partial^2 \tilde{\varphi}_i}{\partial \tilde{z}_i^2}, 
\label{eq:avg_pnp}
\end{equation}
where $\tilde{\mathcal{D}}_i=\tilde{D}\kappa_i I_0(\kappa_i)/(2I_1(\kappa_i))$ is the effective charge diffusivity in the $i$-th pore and $I_n$ is the modified Bessel function of the first kind of order $n$. 
For $\kappa_i\gg 1$, i.e., the thin-double-layer regime, $\tilde{\mathcal{D}}_i \sim \tilde{D}\kappa_i/2$ and the diffusion coefficient is set by electromigration alone. For the overlapping double layer regime, where $\kappa_i\ll 1$ and $\tilde{\mathcal{D}}_i \sim \tilde{D}$, the diffusion coefficient is set by diffusion alone. For intermediate $\kappa_i$, the diffusion coefficient is set by a balance of electromigration and diffusion. This balance could be responsible for a slight increase in experimentally measured in-pore diffusivity for a negative polarization \cite{forse2017direct}, since in this setup the counterion adsorption and co-ion desorption were noted to be equal, such that Debye-Hückel theory is more likely to give a reasonable qualitative picture.

We stress that $\tilde{\mathcal{D}}_i$ is different in different pores. By controlling the rate of charge transport, $\tilde{\mathcal{D}}_i$ sets the rate of local charge accumulation, such that it is inversely proportional to the pore capacitance per unit volume $\tilde{D}\varepsilon/(\tilde{\mathcal{D}}_i\tilde{\lambda}_D^2)$; see Fig. \ref{fig:schematic}c. The relationship described in \eqref{eq:avg_pnp} for arbitrary $\kappa_i$ was first recognized in our work for a single pore \cite{henrique2022charging}, where it was reported in terms of the average electric potential.

Next, we focus on junction boundary conditions, which are a crucial element of this work. We write the current conservation across each junction as
\begin{subequations}
\label{eq:modkirchoff}
\begin{equation}
    \sum_{i \in \textrm{junction}} \tilde{A}_i \hat{\boldsymbol{n}}_i\cdot\tilde{\nabla}_i\tilde{\varphi}_i|_{z_i(\textrm{junction})} = 0,
    \label{eq:kcl}
\end{equation}
where $\tilde{A}_i$ is the cross-sectional area of a pore, $\hat{\boldsymbol{n}}_i$ is the unit normal vector to $\tilde{A}_i$ pointing away from the junction, and the sum is across all the pores connected to the junction.

Physically, \eqref{eq:kcl} is essentially Kirchhoff's current law, for the identical conductivities of the electrochemical potentials of charge of all pores. However, \eqref{eq:kcl} only provides a single equation relating the potentials $\tilde{\varphi}_i$ across a junction. The next set of equations states the continuity of the electrochemical potential of charge across any junction:
\begin{equation}
    \left. \tilde{\varphi}_i \right|_{z_i(\textrm{junction})} = \left. \tilde{\varphi}_j \right|_{z_j(\textrm{junction})}, \ (i,j) \in \textrm{junction}, \label{eq:kvl}
\end{equation}
\end{subequations}
\noindent where $i$ and $j$ are any distinct pores connected to a junction, while $\tilde{z}_i(\textrm{junction})$ and $\tilde{z}_j(\textrm{junction})$ are the axial coordinates of the points where their centerlines intersect the junction; see Fig. S1b. This boundary condition is equivalent to the continuity of voltage across nodes of electrical circuits. We reiterate that for confined electrolytes \eqref{eq:kvl} cannot be written for the average electric potentials since $\bar{\tilde{\phi}}_i$ could experience a sharp change across a junction. For a circuit represented in terms of $\tilde{\varphi}_i$, \eqref{eq:kvl} enables one to write that the drop of electrochemical potentials of charge across a loop is zero, i.e., $\sum_{i \in \textrm{loop}} \Delta \tilde{\varphi}_i=0$ where $\Delta \tilde{\varphi}_i=\tilde{\varphi}_i(\tilde{z}_i=\tilde{\ell}_i)-\tilde{\varphi}_i(\tilde{z}_i=0)$, or a modified Kirchhoff's voltage law. As such, \eqref{eq:modkirchoff} represent a set of modified Kirchhoff's laws in terms of $\tilde{\varphi}_i$, which enable us to close the system of equations in an arbitrary network of pores; see Fig. \ref{fig:schematic}b,c. We highlight that $\tilde{\varphi}_i$ provides the ability to capture both electromigrative and diffusive fluxes in the system and thus expands the ability to study ionic transport in confined geometries.

We note that to solve \eqref{eq:avg_pnp} in each pore, in addition to \eqref{eq:modkirchoff} as boundary conditions across junctions, we also need initial conditions as well as the boundary conditions at the SDL/pore interface and the dead-ends. These conditions are straightforward and are omitted here for brevity; they are listed in the Methods section and systematically described in the Supporting Information. 

\begin{figure*}[t!]
    \centering
    \includegraphics[width=\textwidth]{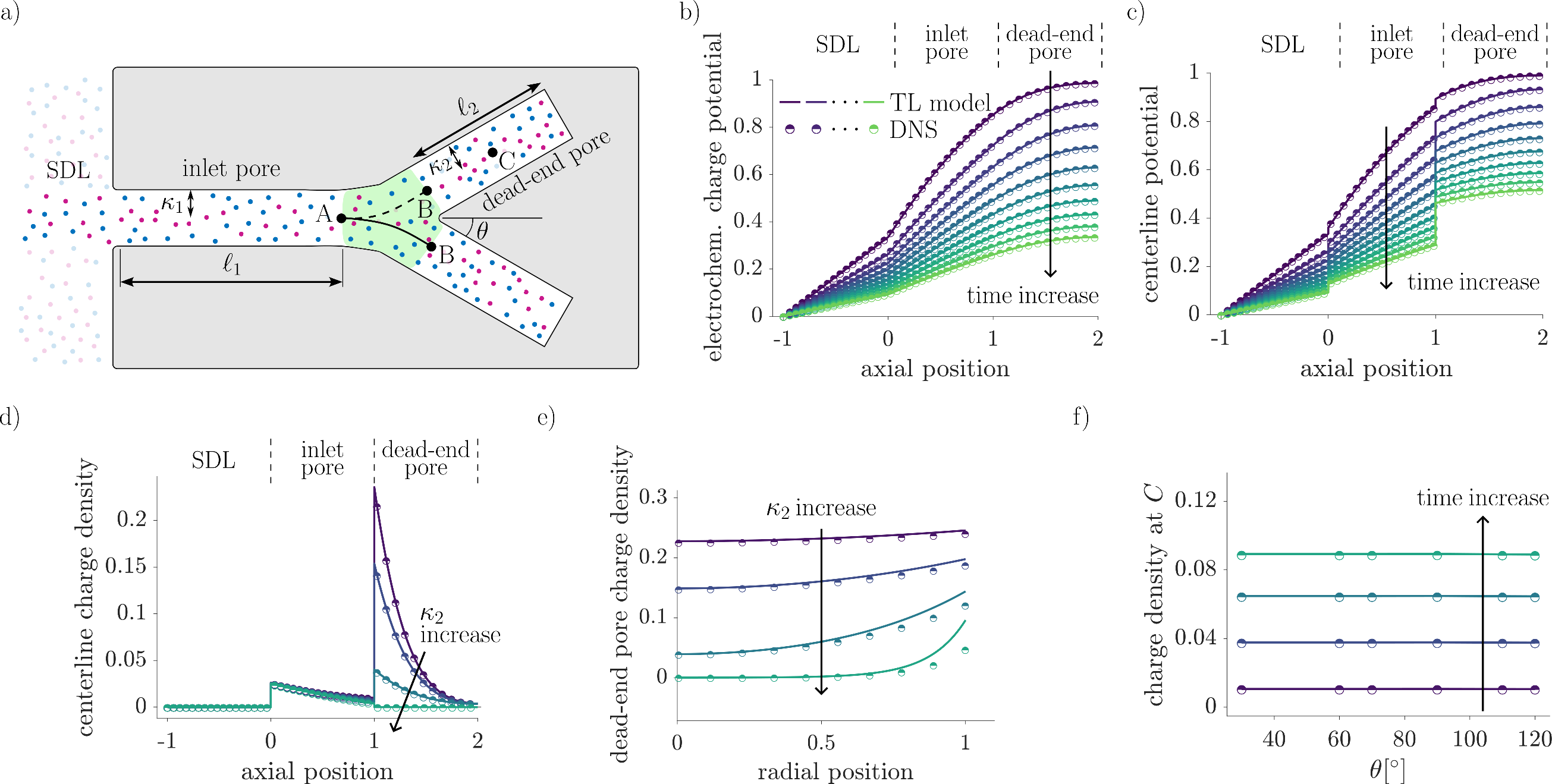}
    \caption{ \textbf{Validation of the TL model for a Y-junction}. Dimensionless properties are defined by $r_i=\tilde{r}_i/\tilde{a}_i$, $z_i=\tilde{z}_i/\tilde{\ell}$, $t=\tilde{t}/(\tilde{\ell}^2/\tilde{D})$, $\phi_i=\tilde{\phi}_i/\tilde{\phi}_D$, $\rho_i=\tilde{\rho}_i/(\varepsilon\tilde{\phi}_D/\tilde{\lambda}_D^2)$, and $\varphi_i=(\tilde{\mu}_+-\tilde{\mu}_-)/(2e\tilde{\phi}_D)$ a) Schematic of the Y-junction geometry of slit pores used for the comparison of the TL model with DNS. The junction is shown in green. b) Electrochemical potential of charge $\varphi_i$ and c) Centerline potential $\phi_i(r_i=0)$ vs. axial positions $z_i$ along the pores and the SDL for ten equally spaced times in the interval $t\in [0.008,0.098]$, with $\kappa_2=2$. TL analytical results are represented by solid lines and DNS results by half-filled orbs for all plots in parts b)--f). The use of effective Kirchhoff's laws for the junction gives accurate results up to points on the order of a width away from the intersection of the centerlines (see also Fig. S3), with good matching with the DNS over all times examined. d) Centerline charge density vs. axial position along the pores and the SDL and e) Dead-end pore charge density at $z_2=0.1$, $\rho_2(z_2=0.1)$ vs radial coordinate $r_2$, both for all dead-end pore relative sizes $\kappa_2\in\{0.4,0.8,2,8\}$. The TL model presents good agreement with DNS even at $z_2=0.1$, about $100$ nm from the transition region. For higher pore sizes, with larger geometrical transition region widths, some disagreement may be found in the double layers close to the junction. f) Charge density at the centerline midpoint $C$ of the dead-end pore, $\rho_2(r_2=0,z_2=0.5,t)$, vs. the half-angle between the centerlines of the dead-end pores, $\theta$, for four equally spaced times $t\in [0.01,0.04]$. The TL model prediction of $\theta$ dependence of the independent variables in long pores is supported by the DNS.}
    \label{fig:y_junction}
\end{figure*}

\subsection{Validation}

To probe the accuracy of the proposed asymptotic model, we perform direct numerical simulations (DNS) of the full PNP equations for a simple geometry -- the Y-junction shown in Fig. \ref{fig:y_junction}a. There, an inlet slit pore is connected on one end to the reservoir through an SDL and on the other end to two identical dead-end slit pores. We note that the slit-pore geometry was adopted to facilitate the DNS; the Supporting Information details the minor modification of the framework that the slit-pore geometry requires. Figs. \ref{fig:y_junction}b and \ref{fig:y_junction}c compare the dimensionless electrochemical potential of charge $\varphi_i(z_i, t)$ and centerline potential  $\phi_i(r_i=0,z_i,t)$ as predicted by the TL model and the DNS. We note that there is an excellent agreement for both the early and late dynamics. Fig. \ref{fig:y_junction}b, c demonstrate that while $\varphi_i$ is continuous across the SDL/pore interface as well as the junction, $\phi_i(r_i=0)$ has sharp changes; see also Fig. S3. Note that these sharp changes are also present at steady-state since double layers are screened to different extents according to the relative pore sizes.

We compare the value of charge profiles in Fig. \ref{fig:y_junction}d--f. Fig. \ref{fig:y_junction}d shows that the theory predicts the centerline charge density profile adequately from $\kappa_2=0.4$ to $\kappa_2=8$, a range that extends from the overlapping- to the thin-double-layer limit. We find a very good agreement between the proposed model and the DNS.  To further understand the dependency of the matching on the junction characteristics, we show the dead-end pore charge profile vs. the radial coordinate in Fig. \ref{fig:y_junction}e at $z_2=0.1$, dimensionally a 100 nm away from the intersection of centerlines of the inlet and dead-end pores. We note that there is good agreement between the theoretical profiles and DNS for moderately to highly overlapping double layers. For larger dead-end pore sizes, the length scale of the influence of the junction region on the charging profile becomes more pronounced, such that a slight quantitative disagreement is noticed in the thin double layers. 

A key inference of the proposed model is the independence of the charge and potential profiles on the angles between long pores in a junction. In fact, in the assumed asymptotic regime of slender pores, transport becomes one-dimensional and the split of current at any junction becomes only a function of the two properties: the number of connected pores and their radii, as seen from Kirchhoff's current law -- \eqref{eq:kcl}. We test this hypothesis by comparing the time evolution of the charge density at the centerline midpoint of the dead-end pores for multiple angles between the centerlines of these pores. Fig. \ref{fig:y_junction}f shows that, as predicted by the model, the charge profile is nearly independent of this angle over time. This result is valid across a wide range of angles, from $2 \theta = 60 ^{\circ}$ to $2 \theta = 240 ^{\circ}$. It implies that our results should hold for pore connections in three-dimensional networks, provided the pores are slender.

The above analysis demonstrates that the proposed approach based on effective Kirchhoff's laws in terms of $\varphi_i$ is able to recover all the crucial effects of the Y-junction to the charging process. It should be noted that this approach applies to arbitrary networks of long pores and does not require that the electric double layers be thin, or that the porous structure be periodic, and it is computationally affordable. It is 5 orders of magnitude faster than the DNS and thus is highly scalable. In the remainder of the article, we focus on the model insights of the model on pore connectivity and spatial arrangement effects. To this end, we undertake case studies of simple lattice geometries, though the model can be applied to three-dimensional pore networks.  

\subsection{Effects of the Spatial Arrangement of Pores}

\begin{figure*}[ht!]
    \includegraphics[width=\textwidth]{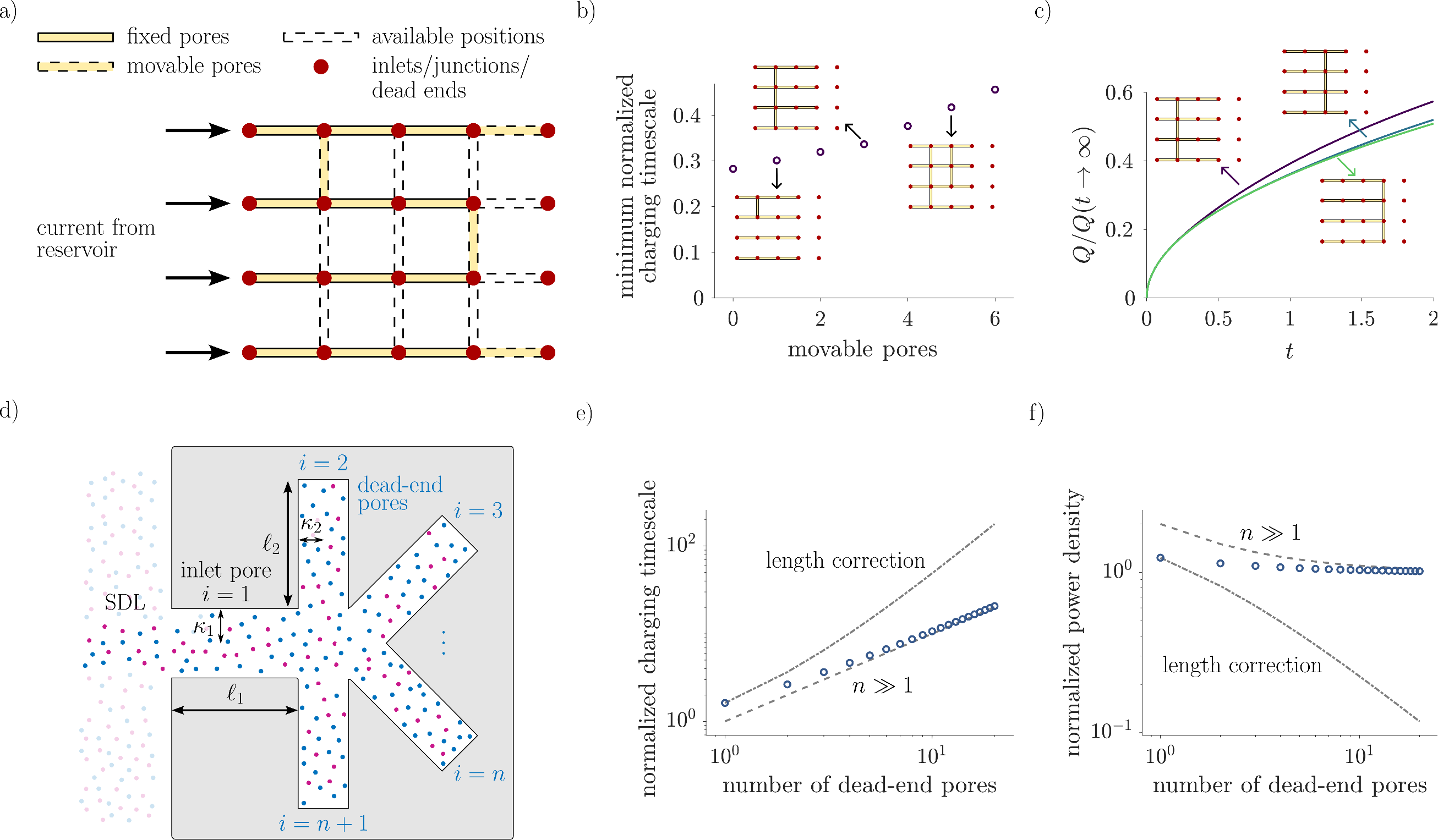}
    \caption{\textbf{Effects of spatial arrangement on the charging timescale of a lattice.} a) Schematic of the possible pore arrangements on the $4\times 5$ lattice. Pores are shown as light yellow rectangles; inlets, junctions, and dead ends are shown as red circles. Current flows from the reservoir into the pores (for $\tilde{\phi}_D<0$). The 12 horizontal pores outlined by solid lines are fixed on the first three columns of each row. $X$ movable pores outlined by dashed lines are allowed to be placed in the positions indicated by dashed rectangles. b) Minimal charging timescale of all possible configurations (normalized by the lattice scale $(3\tilde{\ell})^2/\tilde{D}$; see Methods) vs. the number of movable pores, $X$. Optimal configurations are achieved by placing the pores as close to the reservoir as possible. c) Fraction of the total lattice charge over the steady-state limit vs. time (normalized by $\tilde{\ell}^2/\tilde{D}$) for the configurations shown in the plot, with $X=3$. The smaller length traversed by ions in pores close to the reservoir optimizes late-time charging. Early-time charging is not influenced by these arrangements. \textbf{A toy model for current division vs. length increase effects.} d) Schematic of the junction with $n$ dead-end pores. An inlet pore with relative pore size $\kappa_1$ and dimensionless length $\ell_1$ is connected on one end to an SDL and on the other end to an arbitrary number $n$ of identical dead-end pores with relative pore sizes $\kappa_2$ and length $\ell_2$. e) Normalized charging timescale of the junction $\tau\mathcal{D}_1$ and f) Normalized power density of the junction $\mathcal{P}/A_1$ (both blue orbs) vs. the number of dead-end pores for identical inlet and dead-end pores with negligible SDL charge transfer resistance ($\mathrm{Bi}_1\to\infty$). The limit of large coordination number $n\gg 1$ is shown by the gray dashed line and the result of the estimation of the dynamics by a single pore with the total length of the network, $N=1$ and $\ell_1=n+1$, is shown by the gray dash-dotted line.}
    \label{fig:matchsticks}
\end{figure*}

A limitation of the existing theoretical frameworks for EDL charging in porous media, such as porous-electrode theories \cite{newman1962theoretical,biesheuvel2010nonlinear,mirzadeh2014enhanced}, stack- \cite{lian2020blessing} and laminate-electrode \cite{taocharging} theories, is the inability to resolve microstructural features of arbitrary porous structures, such as coordination numbers and distinct arrangements of pore sizes. To illustrate the application of the current TL model to the study of such effects, we first consider the porous network shown in Fig. \ref{fig:matchsticks}a.  Current flows from the reservoir into the pores through SDLs for $\tilde{\phi}_D<0$. The network consists of a $4\times 5$ array of nodes, representing electrode inlets, junctions, or pore dead ends. These nodes may be connected by horizontal or vertical pores, represented by light yellow rectangles. To focus on the characterization of distinct configurations at the microscale, we compare structures with the same number of pores, but in different positions of the lattice. We fix 12 horizontal pores in the first three columns of all four rows and let an arbitrary number $X$ of pores be moved. They may be placed in any of the positions outlined by dashed lines in Fig. \ref{fig:matchsticks}a. The rules of pore placement resemble a matchsticks puzzle, but the question here is: what spatial arrangement of pores achieves the lowest charging timescale?

For each number of movable pores, $X$, there is at least one configuration with a minimal charging timescale. These minimal charging timescales are plotted as a function of the number of movable pores in Fig. \ref{fig:matchsticks}b, along with the optimal configurations. We see that the strategy for achieving optimal charging consists in placing the mobile pores close to the reservoir, such that the length traversed by the ions to reach their equilibrium positions is reduced. In fact, Fig. \ref{fig:matchsticks}c shows that moving vertical connections away from the reservoir does not affect the early stage of the charging process, but it decelerates the late-time dynamics. Perhaps surprisingly, placing the mobile pores near the reservoir is the optimal strategy despite causing early divisions of the current at the vertical connections. One available option would be to place the mobile pores horizontally after the fixed pores, making an effectively longer capillary bundle. Fig. S4 shows a roughly linear increase in the average charging timescale with the number of vertical pores, for all consistent configurations, like a stack electrode \cite[Eq. 4]{lian2020blessing}. On the other hand, porous-electrode theory predicts increases with the number of pores squared \cite[Eq. 20]{biesheuvel2010nonlinear}. The advantage of the present model is its ability to map the microstructure to a charging timescale, with dependencies on the number of pores which may vary from the stack-electrode to the porous-electrode extremes.

The competing effects of length increase and current division are better understood from a simple toy-model geometry: we examine theoretically a junction like the one in Fig. \ref{fig:y_junction}a, but with an arbitrary number $n$ of dead-end pores, as shown in Fig. \ref{fig:matchsticks}d, to control the extent of current division. From the analytical solution developed in the Supporting Information, Fig. \ref{fig:matchsticks}e shows that increases in the coordination number of the connection, $n+1$, 
increase the charging timescale roughly linearly (asymptotically for $n\gg 1$) as a result of the inversely proportional decrease of the current entering each dead-end pore. Since the capacitance increases linearly with the coordination number, Fig. \ref{fig:matchsticks}f shows that the power density is weakly dependent on the coordination number. Nevertheless, for an equivalent single pore with the same total length of the inlet and dead-end pores, both the pore capacitance and resistance increase linearly with length. This results in a quadratic increase of the charging timescale of the single pore with the number of pores that are pieced together to form it and a sharp decrease of the power density, as shown in Figs. \ref{fig:matchsticks}e-f. Thus, connections of pores near the reservoir are preferable to longer capillary bundles.

\begin{figure*}[ht!]
    \centering
    \includegraphics[width=\textwidth]{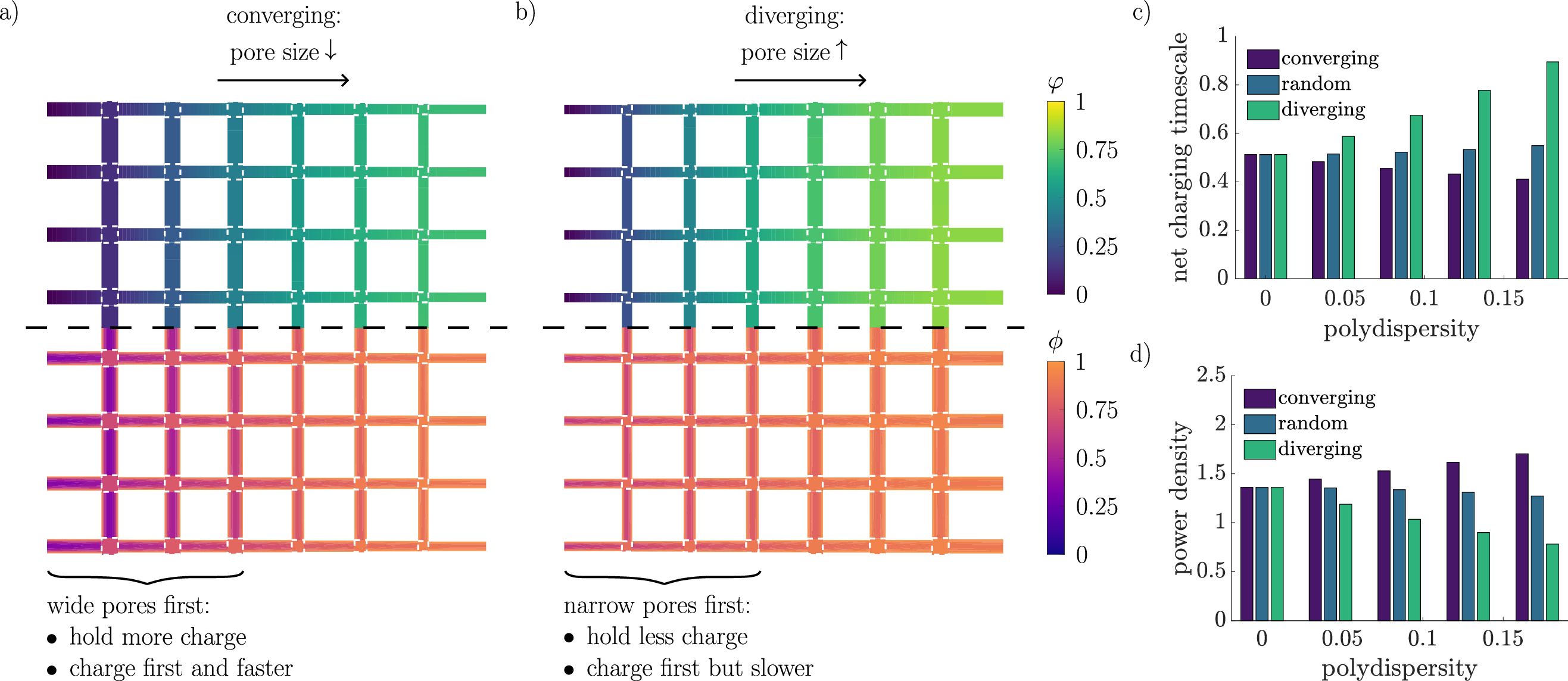}
    \caption{\textbf{Effects of Polydispersity and Configuration in the Charging Characteristics of Lattices.} a) and b) Contour plots of charge density in $8\times 8$ fully connected pore network lattices with different arrangements of a set of relative pore sizes drawn from a log-normal distribution with average relative pore size $\langle\kappa\rangle=2$ and $\mathrm{polydispersity}=\langle\kappa\rangle/\mathrm{sd}(\kappa)=0.17$. Pore sizes are a) decreasing (``converging'') and b) increasing (``diverging'') in the direction of charging. c) Lattice net charging timescale, i.e., the time required to reach 70.18$\%$ of steady-state charge, and d) lattice power density vs. polydispersity for converging, random, and diverging orders of pore sizes in the direction of charging. The importance of the pore size arrangement increases with the polydispersity. For converging configurations, when the largest pores are near the reservoir, their more effective potential screening promotes faster charging. Since they hold the highest fractions of charge and are charged first, the net charging timescale decreases. On the other hand, in a diverging configuration, the diffusion-dominated transport in the inlet pores is a bottleneck. While it does not affect the capacitance, the power density decreases by a factor of 2.2 compared to the previous case.}
    \label{fig:poly}
\end{figure*}

\subsection{Combined Effects of Polydispersity and Spatial Arrangement}

The analysis in the prior section highlights the sensitive dependency of the net charging timescale on the arrangement of the pores. However, the literature primarily uses the pore size distribution to characterize the electrode structure \cite{presser2012electrochemical, zhang2014highly}. This invites the question: given the polydispersity of a pore network, how sensitive is its charging timescale to the arrangement of the pores?

To address this question, we consider $8\times 8$ fully connected pore lattices with pore sizes sampled from log-normal distributions with different polydispersities and arrange them in three different configurations. The first one, ``converging'', is shown in Fig. \ref{fig:poly}a. The horizontal and vertical pore sizes both decrease in the direction of charging. A second one, ``random'', is realized when the placement of pores is determined randomly by the order in which their sizes were drawn. The last one, ``diverging'',  shown in Fig. \ref{fig:poly}b, has increasing pore sizes in the direction of charging. The top and bottom halves of the lattices in Figs. \ref{fig:poly}a and b show, respectively, the contours of the electrochemical potential of charge and the electric potential at a given time. Fig. \ref{fig:poly}a again highlights that the electrochemical potential of charge is predicted to be constant across the junctions, whereas the electric potential changes across junctions to reflect the contrast in steady-state profiles arising from the different extents of electric field screening in pores of unequal sizes.

Qualitatively, Fig. \ref{fig:poly}a and \ref{fig:poly}b show that the profiles of the converging and diverging arrangements of the pores are different, even if the polydispersity is identical.  Specifically, we observe that the converging arrangement charges faster, as further shown in movies S1 and S2. In the converging scenario, the relative pore sizes $\kappa_i$ are larger near the reservoir, which causes the diffusion coefficients $\mathcal{D}_i$ to be larger as they are dominated by electromigration.  In contrast, in the diverging scenario, the narrow inlet pores are bottlenecks; their $\mathcal{D}_i$ are smaller since they are diffusion-dominated. Recent molecular dynamics results have reported a similar strategy of a converging cross-section of a nanometer-wide pore for ion transport optimization \cite{mo2023horn} -- though at this length scale, it is a result of the reduction of steric hindrance of co-ion desorption, not addressed in this work.

Quantitatively, the charging timescale and power density for all the configurations are shown in Fig. \ref{fig:poly}c and \ref{fig:poly}d. We find that for higher polydispersities, the difference between the three configurations becomes more pronounced. For a moderate polydispersity of 0.17, the relative pore sizes of the largest and smallest horizontal pores were 2.35 and 1.18, differing by a factor of two. In this case, the charging timescales of the converging and diverging timescales differ by a factor of 2.2. Even when compared to the random lattice, the converging configuration showed a power density 1.3 times higher. Interestingly, higher polydispersity may be even desirable for the converging scenario since inlet pores significantly influence the charging timescale. However, such a preference is only likely to hold for a low porosity system since volume constraints, which are currently not accounted for in the aforementioned results, will start to impact the analysis. 

Crucially, our results highlight that polydispersity alone is not sufficient to understand the charging and discharging dynamics of an electrode, and more consideration should be given to parameters describing the connectivity and spatial arrangement of pores. These particular effects and analyses are made possible due to the theoretical framework outlined in this manuscript and, as such, had not been explored in prior literature.

\section{Conclusions}
We propose a model for the electric-double-layer charging in arbitrary networks of long pores which demonstrates effective Kirchhoff's laws based on the electrochemical potential of charge, defined here as the valence-weighted average of the ion electrochemical potentials $\tilde{\varphi}=(\tilde{\mu}_+-\tilde{\mu}_-)/(2e)$. The proposed methodology is able to recover the spatial and temporal dependencies of charge density and electric potential obtained from direct numerical simulations but with a speed that is six orders of magnitude faster. We briefly discuss the implications of the model on the effects of pore arrangement and polydispersity and uncover the interplay of these factors on the dynamics of electrode charging for idealized pore networks.

Our work has broad implications for the characterization of ionic transport in porous media. For instance, our work provides a framework to explore the impact of pore network morphology through connectivity, tortuosity, and polydispersity simultaneously to connect the microstructure to macroscopic properties, which is a crucial knowledge gap in the literature \cite{lian2020blessing,pedersen2023equivalent}. One of the promising avenues of our work is its impact on electrode impedance spectroscopy and mapping out effective circuits for different porous networks. The electrochemical experiments rely significantly on EIS but do not currently have a methodology to map out the impact of connectivity, apart from the single-pore TL models. As an example, one can take a 3D microstructure of an electrode, convert it into a ball-stick model, and employ our methodology to forecast Nyquist plots of impedance. Since we do not apply restrictions on double-layer thickness, this can provide insights into the charging dynamics that were not possible previously. 

The work also opens up opportunities for incorporating the effects of the geometry of the pores and other interaction potentials, which have been argued to be important under confinement. While the analysis above is limited to the classical treatment of linearized Poisson-Boltzmann systems, our framework is general and the idea of employing the equality of the electrochemical potential of charge to evaluate electric potential jumps across a junction could be extended to include other interaction parameters. Other effects such as diffusivity contrast \cite{henrique2022impact}, mixtures of electrolytes \cite{gupta2018electrical, jarvey2022ion}, and surface reactions \cite{biesheuvel2011diffuse, jarvey2022ion, jarvey2023asymmetric} are also possible to incorporate.

From a practical standpoint, the framework presents a rational methodology to design 3D-printed electrodes \cite{fu2016graphene}, which are gaining traction in the literature, especially for low-tortuosity materials. The analysis of charging dynamics for a given 3D-printed electrode structure could be carried out through this approach and it could thus guide the design of electrodes. Additionally, there is a growing interest in developing electrodes with pseudocapacitive materials, where there is an interplay of EDL charging in porous media and surface reactions, which our methodology can handle by changing the ideally blocking electrode condition to a reactive flux condition \cite{simon2008materials, simon2020perspectives, jarvey2022ion, jarvey2023asymmetric}. Such a complete continuum model of pseudocapacitive charging of porous media would be fundamental as a basis with which to compare results of experiments and molecular dynamics simulations \cite{fleischmann2022continuous}, allowing studies to parse apart intrinsically molecular effects from the predictions of a classical Debye-Hückel theory for reactive porous media.

\section*{Methods}
\subsection*{Solution of the transmission-line model}

The dimensionless form of the transmission line equation, \eqref{eq:avg_pnp}, and the boundary conditions in \eqref{eq:modkirchoff}, is supplemented by the initial condition
\begin{equation}
    \varphi_i(z_i,t=0)=1
\end{equation}
for all pores, the inlet boundary conditions
\begin{subequations}
\begin{equation}
    \left.\dfrac{\partial\varphi_i}{\partial z_i}\right|_{z_i=0}=\mathrm{Bi}_i\varphi_i(z_i=0,t)=1
\end{equation}
for pores connected to SDLs, where $\mathrm{Bi}_i=A_{s,i}/(A_i\ell_{s,i})$ is a charge transport Biot number analogue of the $i$-th inlet pore, and the no-flux boundary conditions 
\begin{equation}
    \left.\dfrac{\partial \varphi_i}{\partial z_i}\right|_{z_i=\ell_i}=0
\end{equation}
\label{eq:bcs}
\end{subequations}
for dead ends of pores. More details about the initial and boundary conditions are provided in the Supporting Information. 

We solve this system of equations for a single junction with $n$ dead-end pores analytically in the Supporting Information. Though the analytical method is also applicable to general networks, it becomes more cumbersome. Therefore, we solve the equations for lattices numerically. The governing equations are discretized only in space by second-order central finite differences. Labeling the points by the superscript $j$, the $m$ grid points in the $i$-th pore are $z_i^j=(j-1)\Delta z_i$ for $j=1,\cdots,m$, where the mesh width is $\Delta z_i=\ell_i/(m-1)$. The grid function $\varphi_i^1$, $\varphi_i^2$, $\cdots$, $\varphi_i^m$ consists of the approximations of the electrochemical potential of charge at the grid points, $\varphi_i(z_i^j,t)\approx \varphi_i^j(t)$. In terms of this grid function, second derivatives in the axial coordinate are approximated in the inner points by
\begin{subequations}    
\begin{equation}
    \left.\dfrac{\partial^2\varphi_i}{\partial z_i^2}\right|_{z_i=z_i^j}\approx \dfrac{\varphi_i^{j-1}-2\varphi_i^{j}+\varphi_i^{j+1}}{(\Delta z_i)^2},\quad j=2,\cdots,m-1
\end{equation}
and second-order one-sided approximations are used for the first derivatives at the boundary points, namely
\begin{equation}
    \left.\dfrac{\partial\varphi_i}{\partial z_i}\right|_{z_i=0}\approx -\dfrac{3\varphi_i^1-4\varphi_i^2+\varphi_i^3}{2\Delta z_i}
\end{equation}
and
\begin{equation}
    \left.\dfrac{\partial\varphi_i}{\partial z_i}\right|_{z_i=\ell_i}\approx
    \dfrac{\varphi_i^{m-2}-4\varphi_i^{m-1}+3\varphi_i^{m}}{2\Delta z_i}.
\end{equation}
\label{eq:fd_approx}
\end{subequations}

The system of ordinary differential equations resulting from the discretization of \eqref{eq:avg_pnp} and the boundary conditions in \eqref{eq:modkirchoff} and \eqref{eq:bcs} couples the discretized electrochemical potentials of charge of all pores. To solve it, we treat the boundary conditions as algebraic constraints and use \texttt{ode15s}, MATLAB's differential algebraic equation solver, sampled at a time array logarithmically spaced in the interval $[10^{-6},2]\times (N_\ell-1)^2$, where $N_\ell$ is the number of pores per side on the lattice. We choose a number $m=50$ of grid points in each pore to achieve grid-size independence of the total current going into the networks (to within 1\% of the current for $m=100$) and retrieve the correct total network charge, known from the steady-state charge density $\bar{\rho}_i\to -1/\mathcal{D}_i$; see Supporting Information. 

\subsection*{Pore Network Plots and Properties}

DNS in Fig. \ref{fig:y_junction} are performed with the parameters $\tilde{c}_\infty=1$ mM, $\tilde{\phi}_D=10$ mV, $\tilde{\ell}=\tilde{\ell}_1=\tilde{\ell}_2=\tilde{\ell}_3=1\mu$m, $\tilde{D}=1.34\times 10^{-9}\mathrm{m}^2/s$ and $kT/e=25.7$ mV. The inlet pore has a relative pore size (half-width by Debye length) $\kappa_1=4$ and is connected to the SDL with a Biot number $\textrm{Bi}_1=2$.

All pore network simulations in Figs. \ref{fig:matchsticks}, \ref{fig:poly} and S3 are performed in the limit $\mathrm{Bi}_i\to\infty$. Contour plots of the electrochemical potential of charge and the electric potential are respectively colored using viridis and plasma, perceptually uniform colormaps. The junctions, not resolved by the model, are delineated by white dashed squares, and colored according to the arithmetic average of the properties on all sides. This is done to represent that current flows through the junctions as well. The pore sizes of the polydisperse lattices were drawn from pseudorandom samples of log-normal distributions using MATLAB's \texttt{lognrnd} routine, which takes as inputs the parameters $\mu$ and $\sigma$ of the distribution, calculated based on the desired mean and variance.


The net charging timescale $\tau_\mathrm{num}$ is numerically calculated in Figs. \ref{fig:matchsticks}b, \ref{fig:poly} and S3 as the time required for the network to reach 70.18\% of its steady-state charge divided by $(N_y-1)^2$, where $N_y$ is the number of nodes of the lattice in the direction normal to charging, having square lattices where $N_x=N_y$ as the reference. The division by the number of pores per side squared factors out the length of lattice, making lattices of different numbers of pores comparable, and the percentage is chosen such that  $\tau_\mathrm{num}=4/(\pi^2\mathcal{D})$ for capillary bundles \cite{henrique2022charging}. Here, $\mathcal{D}$ is the common effective diffusivity of all pores. On the other hand, the dimensionless extensive capacitance of a lattice is known,
based upon the steady-state charge density $\bar{\rho}_i\to -1/\mathcal{D}_i$ as $t\to\infty$ (which is already normalized by the applied electric potential) to be $C=\sum_{i=1}^N(A_i/\mathcal{D}_i)$ for equal pore lengths. In the plots of Figs.  \ref{fig:poly} and S3, with $N_x=N_y=N_\ell$, we normalize the extensive capacitance by the total volume of the fully connected monodisperse lattice with the same pore lengths and average relative sizes, $2\pi(N_\ell-1)^2\langle\kappa\rangle^2$. Therefore, the capacitance per unit volume that we report is calculated as
\begin{equation}    C=\dfrac{\sum_{i=1}^N(\kappa_i^2/\mathcal{D}_i)}{2(N_\ell-1)^2\langle\kappa\rangle^2},
\label{eq:C_def}
\end{equation}
where $\langle\kappa\rangle$ is the imposed average radius of the log-normal distribution. When multiplied by the porosity of the reference monodisperse configuration, \eqref{eq:C_def} yields the dimensionless electrode volumetric capacitance, i.e., the capacitance divided by the electrode volume.

The power density is simply calculated as the ratio of capacitance to charging timescale,
\begin{equation}
    \mathcal{P}=\dfrac{C}{\tau_\mathrm{num}}.
\end{equation}

\subsection*{Direct Numerical Simulations}
The non-linear PNP equations are
\begin{subequations}
\begin{equation}
    \dfrac{\partial \tilde{c}_\pm}{\partial \tilde{t}}=\tilde{D}\left[\nabla^2\tilde{c}_\pm\pm \dfrac{e}{kT}\nabla\cdot(\tilde{c}_\pm\tilde{\nabla}\tilde{\phi})\right]=-\tilde{\nabla}\cdot\tilde{\mathbf{N}}_\pm
    \label{eq:np_dns}
\end{equation}
where $\tilde{\mathbf{N}}_\pm$ are the dimensional ionic fluxes, and
\begin{equation}
    -\varepsilon\tilde{\nabla}^2\tilde{\phi}=e(\tilde{c}_+-\tilde{c}_-).
    \label{eq:poisson_dns}
\end{equation}
\label{eq:pnp_dns}
\end{subequations}
These equations were solved numerically by direct numerical simulations for the Cartesian Y-junction geometry illustrated in Fig. \ref{fig:y_junction}a. As shown in Fig. S2, only half of the domain is represented in the simulations due to top-down symmetry about the colinear centerlines of the SDL and the inlet pore. The domain consists of the SDL, with a centerline of length $\tilde{\ell}_{s}=1$ $\mu$m and a half-width of $76$ nm, decreased by a circular fillet of radius $38$ nm at the connection to the inlet pore, of length $\tilde{\ell}_{1}=1$ $\mu$m and half-width $38$ nm. The end of the inlet pore is then connected to a dead-end pore of centerline length $\tilde{\ell}_{2}=1$ $\mu$m and variable half-width; see Fig. \ref{fig:y_junction}b--f. This junction has a fillet of an eighth of a circle of radius $38$ nm at its right side.

The finite-volume method \cite{eymard2000finite,versteeg2007introduction} was employed for the solution using the open-source software OpenFOAM \cite{weller1998tensorial,jasak2007openfoam}. We describe the domains of application of the initial and boundary conditions in the terms indicated in Fig. S2. To this end, first \eqref{eq:poisson_dns} alone was run under electroneutral conditions to set up the electric field in the static diffusion layer, producing the initial conditions: $\tilde{\phi}$ a solution of
\begin{subequations}
\begin{equation}
\tilde{\nabla}^2\tilde{\phi}=0
\end{equation}
and
\begin{equation}
\tilde{c}_\pm=\tilde{c}_\infty
\end{equation}
\end{subequations}
in the entire electrolyte domain. The boundary conditions are the known electric potential and ion concentrations 
\begin{subequations}    
\begin{equation}
\tilde{\phi}=0
\end{equation}
and
\begin{equation}
\tilde{c}_\pm=\tilde{c}_\infty
\end{equation}
on the interface with the reservoir -- the left side of SDL, either symmetry or no-flux and no electric field conditions
\begin{equation}
\hat{\boldsymbol{n}}\cdot\tilde{\nabla}\tilde{\phi}=0
\end{equation}
and
\begin{equation}
\hat{\boldsymbol{n}}\cdot\tilde{\nabla}\tilde{c}_\pm=0
\end{equation}
on all centerlines (of both the SDL and the pores) and the bottom surface of the SDL, the no-flux and known applied potential (galvanostatic condition) on the bottom surface of the inlet pore and both surfaces of the simulated dead-end pore,
\begin{equation}
\tilde{\phi}=\tilde{\phi}_D
\end{equation}
and
\begin{equation}
\hat{\boldsymbol{n}}\cdot\tilde{\nabla}\tilde{\mathbf{N}}_\pm=0.
\end{equation}
\label{eq:bcs_dns}
\end{subequations}
The numerical solution of \eqref{eq:pnp_dns}--\eqref{eq:bcs_dns}, performed using a 28-core workstation made available by the Princeton Research Computing resources, had a computational cost of 4 million seconds for 1 second of time elapsed in the simulation.

\paragraph{Acknowledgements} 
 
The authors would like to acknowledge the helpful input provided by Gesse Roure, Nitish Govindarajan, Tiras Lin, Howard Stone, and others. The authors acknowledge that the simulations reported in this contribution were performed using the Princeton Research Computing resources at Princeton University which is a consortium of groups including the Princeton Institute for Computational Science and Engineering and the Princeton University Office of Information Technology’s Research Computing department.  A.G. thanks the National Science Foundation (CBET - 2238412) CAREER award for financial support. F.H. thanks the Ryland Graduate Family Fellowship for financial assistance. P.J.Z. would like to acknowledge the support of a project that has received funding from the European Union's Horizon 2020 research and innovation program under the Marie Skłodowska-Curie grant agreement No. 847413 and was part of an international co-financed project founded from the program of the Minister of Science and Higher Education entitled ``PMW" in the years 2020–2024; agreement No. 5005/H2020-MSCA-COFUND/2019/2

\paragraph{Author Contributions}

F. H., P. J. Z., and A. G. designed research; F. H. and P. J. Z. performed research; F. H. and A.G. analyzed data; F. H., P. J. Z., and A. G. wrote the paper.

\paragraph{Competing Interests} The authors declare no conflict of interest.

\setcounter{figure}{0}
\renewcommand{\figurename}{Fig.}
\renewcommand{\thefigure}{S\arabic{figure}}	
\setcounter{section}{0}
\setcounter{equation}{0}

\title{Supplementary Material: Modified Kirchhoff's Laws for Electric-Double-Layer Charging in Porous Media}
\maketitle

In Section 1 we provide details of the derivation of the transmission line model for ion transport in binary symmetric electrolytes confined in networks of cylindrical or slit pores with arbitrary Debye lengths. In Section 2 we describe an analytical method of solution of the model for arbitrary pore networks. We demonstrate its usage to determine electrolyte charging in a simple geometry, namely, a network with a single inlet pore connected on one end to a static diffusion layer and on the other end to an arbitrary number $n$ of 
dead-end pores. Then, we highlight some physical insights of the method. This solution is used to validate the model against direct numerical simulations of the full Poisson-Nernst-Planck in the main text. In Section 3, we comment on aspects of the numerical methods used to solve the transmission line equations. We illustrate the pore configuration dependency of the charging timescale in movies 1 and 2.

\section{Mathematical Modeling}

We develop a pore-scale model for slender pores with cylindrical or slit-shaped cross-sections. The mathematical model is a generalization of the single-pore asymptotic analysis of Refs. \cite{gupta2020charging,henrique2022charging} to pore networks. Pores are indexed integer variables, e.g., $i$, ranging from 1 to the total number of pores $N$. For cylindrical pores, the slenderness requirement takes the form $\tilde{a}_i\ll\tilde{\ell}_i$, where $\tilde{a}_i$ is the radius and $\tilde{\ell}_i$ the length of the $i$-th pore. Correspondingly, in a network of slit pores, we require $\tilde{h}_i\ll\tilde{\ell}_i$, where $\tilde{h}_i$ are the half-widths of the pores.

\subsection{Dimensionless Governing Equations}

The Poisson-Nernst-Planck (PNP) equations are used to describe the transport of the dilute solute in the entire fluid domain, comprising the ion reservoir, static diffusion layers, and the electrolyte confined by the pores. Using tildes for dimensional variables, these equations take the form \cite{deen1998analysis}
\begin{subequations}
\begin{equation}
    \dfrac{\partial \tilde{c}_\pm}{\partial \tilde{t}}=\tilde{\nabla}\cdot\left(\dfrac{\tilde{D}\tilde{c}_\pm}{k_BT}\tilde{\nabla}\tilde{\mu}_{\pm}\right),
    \label{eq:np}
\end{equation}
\begin{equation}
    \tilde{\mu}_\pm=k_BT\ln(\tilde{c}_\pm)\pm e\tilde{\phi},
    \label{eq:mu_pm}
\end{equation}
\begin{equation}
    -\varepsilon\tilde{\nabla}^2\tilde{\phi}=e(\tilde{c}_+-\tilde{c}_-),
    \label{eq:poisson}
\end{equation}
\end{subequations}
where $\tilde{c}_\pm$ are the cation and anion concentrations, respectively, $\tilde{\mu}_\pm$ their electrochemical potentials, $\tilde{\phi}$ is the electric potential, $k_BT/e$ is the thermal voltage, and $\varepsilon$ is the permittivity of the electrolyte. $\tilde{D}$ is the ionic diffusivity, assumed to be the same for both ions and uniform throughout the electrolyte. These equations can be linearly combined to provide relations for the evolution of the charge density $\tilde{\rho}=e(\tilde{c}_+-\tilde{c}_-)$ in the electrolyte,
\begin{subequations}
\begin{equation}
    \dfrac{\partial \tilde{\rho}}{\partial \tilde{t}}=\dfrac{\tilde{D}}{2k_BT}\tilde{\nabla}\cdot\left[e\tilde{c}\tilde{\nabla}(\tilde{\mu}_+-\tilde{\mu}_-)+\tilde{\rho}\tilde{\nabla}(\tilde{\mu}_++\tilde{\mu}_-)\right],
    \label{eq:np_charge}
\end{equation}
\begin{equation}
    -\varepsilon\tilde{\nabla}^2\tilde{\phi}=\tilde{\rho},
    \label{eq:poisson_charge}
\end{equation}
\label{eq:pnp_charge}
\end{subequations}
where 
$\tilde{c}=\tilde{c}_++\tilde{c}_-$ is the total concentration of ions, i.e., salt density. To develop an analytically tractable reduced-order, we can linearize \eqref{eq:np_charge} in the limit of low applied potentials relative to the thermal voltage: $\epsilon=\tilde{\phi}_De/(k_BT)\ll 1$. In the absence of an applied potential, the electrolyte remains in an electroneutral steady state with a total concentration equal to that of the reservoir, $2\tilde{c}_\infty$, i.e., $\tilde{c}=2\tilde{c}_\infty+\mathcal{O}(\epsilon)$, $\tilde{\rho}=\mathcal{O}(\epsilon)$, $\tilde{\phi}=\mathcal{O}(\epsilon)$, and $\tilde{\nabla}\tilde{\mu}_\pm=\mathcal{O}(\epsilon)$. Neglecting terms of $\mathcal{O}(\epsilon^2)$ furnishes the linearized system of equations
\begin{subequations}
\begin{equation}
    \dfrac{\partial \tilde{\rho}}{\partial \tilde{t}}=\dfrac{\tilde{D}e\tilde{c}_{\infty}}{k_BT}\tilde{\nabla}^2(\tilde{\mu}_+-\tilde{\mu}_-),
    \label{eq:lin_np_charge}
\end{equation}
\begin{equation}
    -\varepsilon\tilde{\nabla}^2\tilde{\phi}=\tilde{\rho}.
    \label{eq:lin_poisson_charge}
\end{equation}
\label{eq:lin_dim_pnp}
\end{subequations}
We note that in the linear regime of low applied potentials, charge transport is driven by gradients of a thermodynamic force, namely, the difference of the electrochemical potentials of the cations and anions, $\tilde{\mu}_+-\tilde{\mu}_-$. Note from \eqref{eq:mu_pm} that this thermodynamic force is a multiple of the electric potential corrected by entropic terms related to the ionic concentrations. To represent this effective  potential as a corrected electric potential, we define the electrochemical potential of charge, 
$\tilde{\varphi}$, as $\tilde{\varphi}=(\tilde{\mu}_+-\tilde{\mu}_-)/(2e)$. In the limit of low applied potentials, using \eqref{eq:mu_pm} and the asymptotic expansions $\tilde{c}_\pm=\tilde{c}_\infty+\mathcal{O}(\epsilon)$, $\tilde{\varphi}$ takes the asymptotic form
\begin{equation}
    \tilde{\varphi}=\tilde{\phi}+\dfrac{\tilde{\lambda}_D^2}{\varepsilon}\tilde{\rho},
\end{equation}
where $\tilde{\lambda}_D=\sqrt{\varepsilon k_BT/(2e^2\tilde{c}_\infty)}$ is the Debye length.
The electrochemical potential of charge plays an important role in the circuit representation for networks of arbitrary relative pore sizes, as discussed in the main text.

To nondimensionalize \eqref{eq:lin_dim_pnp}, we take the average pore length $\tilde{\ell}=(\sum_{i=1}^N \tilde{\ell}_i)/N$ as the characteristic lengthscale, i.e., $\ell_i= \tilde{\ell}_i/\tilde{\ell}$. Now, we define the dimensionless variables $t=\tilde{t}/(\tilde{\ell}^2/\tilde{D})$, $\nabla=\tilde{\nabla}/(1/\tilde{\ell})$, $\phi=\tilde{\phi}/\tilde{\phi}_D$,  $c=\tilde{c}/(2\tilde{c}_\infty)$,  $\varphi=\tilde{\varphi}/\tilde{\phi}_D$, and $\rho=\tilde{\rho}/( \varepsilon\tilde{\phi}_D/\tilde{\lambda}_D^2)$. Thus, the dimensionless form of \eqref{eq:lin_dim_pnp} is
\begin{subequations}
\begin{equation}
    \dfrac{\partial\rho}{\partial t}=\nabla^2\varphi,
    \label{eq:nondim_np}
\end{equation}
where
\begin{equation}
    \varphi=\rho+\phi,
    \label{eq:varphi}
\end{equation}
and
\begin{equation}
    -\nabla^2\phi=\left(\dfrac{\tilde{\ell}}{\tilde{\lambda}_D}\right)^2\rho.
    \label{eq:nondim_poisson}
\end{equation}
\label{eq:lin_pnp}
\end{subequations}

It should be noted that the leading-order coefficient of the total ionic concentration is known in advance from the solution in the absence of an applied potential. In this case, the first-order salt density dynamics does not influence the charge dynamics and therefore it does not need to be resolved for the purposes of this study. For electrolytes with asymmetric ionic diffusivities, the salt dynamics is coupled to the charge dynamics and thus both would need to be simultaneously solved \cite{henrique2022impact}.

\subsection{Reduced-Order Transport Equations in the Pores}

Let us denote the restriction of any variable to the $i$-th pore by the pore number as a subscript, e.g., $\rho_i$. We consider the cases of cylindrical and slit pores separately. First, we derive the effective transport equations for cylindrical pores. After this derivation, we briefly discuss the analogous case of slit pores.

In either geometry, we first show that the electrochemical potential of charge is constant over cross-sections, and then we use Poisson's equation to obtain a relation between charge density and electric potential over each cross-section. Finally, we use this relation to average the linearized PNP \eqref{eq:lin_pnp} in each pore and determine a reduced-order equation for a single variable, namely the electrochemical potential of charge.

\subsubsection{Cylindrical Pores}
Assuming axisymmetry due to both the uniform potential in the perfectly conducting electrode and slender aspect ratios of the pores, the profiles are written as functions of only the dimensionless radial coordinate $r_i=\tilde{r}_i/\tilde{a}_i$, axial coordinate $z_i=\tilde{z}_i/\tilde{\ell}$, and time, e.g., $\rho_i(r_i,z_i,t)$. The ranges of the dimensionless coordinates are $r_i\in [0,1]$ and $z_i\in [0,\ell_i]$. Let us define a relative pore size parameter by $\kappa_i=\tilde{a}_i/\tilde{\lambda}_D$. The dimensionless charge flux in the $i$-th pore can be inferred from the right-hand side of \eqref{eq:nondim_np} as $\mathbf{J}_i=-\nabla_i\varphi_i$, where $\nabla_i$ is the dimensionless 
gradient operator calculated in the local cylindrical axisymmetric coordinates $r_i,z_i$. In the long pore limit $\tilde{a}_i\ll\tilde{\ell}_i$, radial equilibrium can be shown to be asymptotically valid (see Refs. \cite{gupta2020charging,henrique2022charging,alizadeh2017multiscale,aslyamov2022relation,aslyamov2022analytical} for related discussions)
\begin{equation}
    \dfrac{1}{r_i}\dfrac{\partial}{\partial r_i}\left(r_i\dfrac{\partial\varphi_i}{\partial r_i}\right)=0.
    \label{eq:flux_ord_0}
\end{equation}
In the absence of surface reactions, blocking electrode boundary conditions hold, implying the nullity of the radial charge flux at $r_i=1$. Integrating \eqref{eq:flux_ord_0}, we find that the radial flux vanishes at any point in a cross-section, consequently
\begin{equation}
    \dfrac{\partial\varphi_i}{\partial r_i}=0.
    \label{eq:rad_flux_eq}
\end{equation}
Let us denote cross-sectional averages of variables by bars, e.g.,
\begin{equation}    \bar{\rho}_i(z_i,t)=2\int_0^1\rho_i(r_i,z_i,t)r_i\,\mathrm{d}r_i.
\end{equation}
\eqref{eq:rad_flux_eq} shows that the electrochemical potential of charge must be independent of the radial coordinate in each pore, and therefore equal to its cross-sectional average. In terms of charge density and electric potential, this yields the relation
\begin{equation}
\varphi_i(z_i,t)=\rho_i(r_i,z_i,t)+\phi_i(r_i,z_i,t)=\bar{\rho}_i(z_i,t)+\bar{\phi}_i(z_i,t)=\bar{\varphi}_i(z_i,t).
\label{eq:rad_eq}
\end{equation}

In a similar fashion to the radial equilibrium relation, \eqref{eq:flux_ord_0}, Poisson's equation in the limit of long pores can be asymptotically simplified to
\begin{equation}
    -\dfrac{1}{r_i}\dfrac{\partial}{\partial r_i}\left(r_i\dfrac{\partial\phi_i}{\partial r_i}\right)=\kappa_i^2\rho_i,
    \label{eq:poisson_r}
\end{equation}
with symmetry around $r_i=0$ and the known dimensionless applied potential $\phi_i(r_i=1)=1$. Using \eqref{eq:rad_eq}, it can be rewritten as a modified Bessel equation of order zero for the charge density,
\begin{equation}
    r_i^2\dfrac{\partial^2 \rho_i}{\partial r_i^2}+r_i\dfrac{\partial^2 \rho_i}{\partial r_i}-\kappa_i^2r_i^2\rho_i=0.
\end{equation}
Its finite solution, written in terms of the average charge density, is given by
\begin{equation}
    \rho_i(r_i,z_i,t)=\mathcal{D}_i\bar{\rho}_i(z_i,t)\dfrac{I_0(\kappa_ir_i)}{I_0(\kappa_i)}
    \label{eq:rho(r)},
\end{equation}
where $I_j$ is the modified Bessel function of the first kind of order $j$ and $\mathcal{D}_i=\kappa_i I_0(\kappa_i)/(2I_1(\kappa_i))$ is the dimensionless effective charge diffusivity of a symmetric electrolyte in the $i$-th pore, with relative size $\kappa_i$. Physically, it results from the size-dependent interplay of diffusion and electromigration, which boosts charge transport in the pores. For large pores, with fully screened double layers, electromigration dominates and sets a linear increase of the effective diffusivity with pore size, $\mathcal{D}_i\sim \kappa_i/2$ for $\kappa_i\gg 1$.

Note that the parameter $1/\mathcal{D}_i$ has been interpreted in the single-pore literature as the dimensionless charging timescale of the electrolyte in an isolated pore with relative size $\kappa_i$ \cite{gupta2020charging,henrique2022charging}. However, since the charging timescale of a single pore is length dependent through the scale $\tilde{\ell}^2/\tilde{D}$, in our current context of a network of pores with potentially different lengths, it makes more sense to use the effective diffusivity $\mathcal{D}_i$.

The known radial dependency allows one to relate the averages of charge density and the electrochemical potential of charge. In fact, using the BC $\phi_i(r_i=1)=1$ and \eqref{eq:rho(r)} to evaluate \eqref{eq:rad_eq} at $r_i=1$, we show that
\begin{equation}
\varphi_i(z_i,t)=1+\mathcal{D}_i\bar{\rho}_i(z_i,t).
\label{eq:rhophi_to_rhobar}
\end{equation}
Lastly, we find effective equations for the transport of the electrochemical potential of charge. Performing cross-sectional averages of the $\mathcal{O}(\tilde{a}_i^2/\tilde{\ell}_i^2)$ equations in the long-pore asymptotic expansions of \eqref{eq:nondim_np}, radial fluxes vanish due to the ideally blocking boundary conditions, and using \eqref{eq:rhophi_to_rhobar} we obtain a diffusion-like equation for the electrochemical potential of charge,
\begin{equation}
\dfrac{\partial\varphi_i}{\partial t}=\mathcal{D}_i\dfrac{\partial^2\varphi_i}{\partial z_i^2}.
\label{Seq:avg_pnp}
\end{equation}

\subsubsection{Slit Pores} 
The two-dimensional profiles, e.g., $\rho_i(y_i,z_i,t)$, are written as functions of the local transversal and longitudinal coordinates, $y_i\in [-1,1]$ and $z_i\in [0,\ell_i]$, which are respectively scaled by $\tilde{h}_i$ and $\tilde{\ell}$. In this case, we define the relative pore size parameter as   $\kappa_i=\tilde{h}_i/\tilde{\lambda}_D$. This choice of Cartesian coordinates is made to keep $z_i$ as the centerline coordinates. The procedure is analogous to the cylindrical geometry case. Transversal equilibrium follows from \eqref{eq:nondim_np} as
\begin{equation}
    \dfrac{\partial^2\varphi_i}{\partial y_i^2}=0.
\end{equation}
Integrating twice in $y_i$, the transversal equilibrium relation
\begin{equation}
\rho_i(y_i,z_i,t)+\phi_i(y_i,z_i,t)=\bar{\rho}_i(z_i,t)+\bar{\phi}_i(z_i,t)=\varphi_i(z_i,t),
\end{equation}
is obtained, where transversal averages are defined by $\bar{f}_i=\int_0^1f_i(y_i,z_i,t)\,\mathrm{d}y_i$ for arbitrary functions $f_i$ using only the top half of the domain due to symmetry. Poisson's equation can be asymptotically approximated by
\begin{equation}
    -\dfrac{\partial^2\phi_i}{\partial y_i^2}=\kappa_i^2\rho_i,
\end{equation}
and solved by
\begin{equation}
    \rho_i(y_i,z_i,t)=\mathcal{D}_i\bar{\rho}_i(z_i,t)\dfrac{\cosh(\kappa_iy_i)}{\cosh(\kappa_i)},
\end{equation}
where now the effective diffusivity of the $i$-th pore is $\mathcal{D}_i=\kappa_i\coth(\kappa_i)$. This interpretation follows from the resulting effective potential transport equation
\begin{equation}
    \dfrac{\partial\varphi_i}{\partial t}=\mathcal{D}_i\dfrac{\partial^2\varphi_i}{\partial z_i^2}.
\label{eq:avg_pnp_slit}
\end{equation}
Notably, the reduced-order equations of cylindrical and slit pores are analogous, differing only by the definition of the effective diffusivity $\mathcal{D}_i$. The higher effective diffusivity (and resulting smaller capacitance) of slit pores compared to cylindrical pores matches the results of Huang. et. al. \cite{huang2008theoretical} and has been attributed to increased ion separation near curved surfaces \cite{wu2022understanding}.

\subsection{Boundary and Initial Conditions}
\label{sec:ics_bcs}

To close the initial- and boundary-value problem set by \eqref{Seq:avg_pnp} -- or equivalently \eqref{eq:avg_pnp_slit} --, we discuss the initial and boundary conditions for the electrochemical potential of charge in this section.

\subsubsection{Connections}

\begin{figure}
    \centering
    \includegraphics[width=\textwidth]{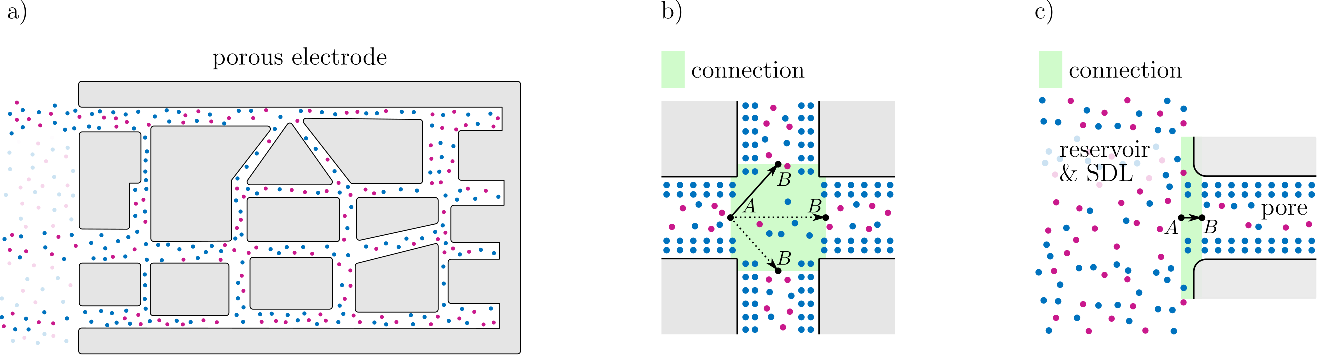}
    \caption{\textbf{Illustration of the treatment of junctions in the model and the paths used to derive boundary conditions for the model}. a) Schematic of a cross-section of the electrode, ions in the electrode represented in purple (cations -- co-ions) and blue (anions -- counterions). b) Zoom-in of a connection between multiple pores and the possible paths $\mathcal{C}_{AB}$ connecting their ends. c) Zoom-in of a pore-SDL connection and the path $\mathcal{C}_{AB}$ connecting their ends.}
    \label{fig:connections}
\end{figure}

The geometry of connections, either between pores or of pores and static diffusion layers, will differ from the assumed cylindrical or slit shapes, as illustrated in Fig. \ref{fig:connections}b,c. However, if these regions are thin -- with a characteristic dimensionless length on the order of the largest relative sizes of the connected pores -- we may neglect such lengths. Thus, we obtain relations between properties of the ends of pores connected by a junction. This procedure also allows us to restrain the analysis to points sufficiently distant from connections, where entrance effects are negligible, such that the long-pore asymptotic expansions and their implication of transversal equilibrium hold.

Properties such as charge density and electric potential may undergo sharp changes over short lengthscales (on the order of the Debye length) and thus they may differ significantly across the ends of connections. These changes are to be expected due to either the unequal profiles promoted by transversal equilibrium in long pores of different relative sizes or due to the electroneutrality of the unconfined static diffusion layers. Such changes have indeed been observed in inlets of isolated pores with overlapping double layers \cite{gupta2020charging} and in pore connections in electroosmotic flows \cite{alizadeh2017multiscale}. In order to determine the appropriate relations of charge density and potential across negligibly thin pore connections, let us consider a path integral of charge flux on the shortest path $\mathcal{C}_{AB}$ - shown in Figs. \ref{fig:connections}b,c - linking any distinct boundary points $A$ and $B$ of a connection, i.e., a transition region. Since the charge flux is conservative in this low-potential model, it can be readily integrated: 
\begin{equation}
\int_{\mathcal{C}_{AB}}\mathbf{J}\cdot\mathrm{d}\boldsymbol{x}=-\int_{\mathcal{C}_{AB}}\nabla\varphi\cdot\mathrm{d}\boldsymbol{x}=\varphi|_{\boldsymbol{x}(B)}^{\boldsymbol{x}(A)},
\label{eq:int_J}
\end{equation}
where $\boldsymbol{x}(A)$ and $\boldsymbol{x}(B)$ are the dimensionless position vectors of the points $A$ and $B$, with $\boldsymbol{x}$ scaled by $\tilde{\ell}$. On the other hand, due to the thinness of the transition region, we may asymptotically neglect the integrals on the LHS of \eqref{eq:int_J}. In fact, parameterizing the path $\mathcal{C}_{AB}$ by the arclength $s$ and using the ML inequality, we estimate the path integral by
\begin{equation}
\bigg|\int_{\mathcal{C}_{AB}}\mathbf{J}\cdot\mathrm{d}\boldsymbol{x}\bigg|=\bigg|\int_{s(A)}^{s(B)}\mathbf{J}\cdot\dfrac{d\boldsymbol{x}}{ds}\mathrm{d}s\bigg|\le \,\ell_{AB}\max_{\boldsymbol{x}\in\mathcal{C}_{AB}}|\mathbf{J}|.
\label{eq:char_flux}
\end{equation}
Now, let us denote the largest pore size in the network by $\tilde{a}=\max_{i\in\{1,\cdots, N\}}\tilde{a}_i$. We assume that the dimensionless charge fluxes are $\max_{\boldsymbol{x}\in\mathcal{C}_{AB}}|\mathbf{J}|=\mathcal{O}(1)$ for $t=\mathcal{O}(1)$ due to the weak applied electric fields, similar to an argument of Chu and Bazant \cite{chu2007surface} for the surface conservation equations of thin EDLs adjacent to blocking surfaces. We also assume that the length of any shortest path connecting endpoints of a transition region is on the order of the largest pore size in the network, $\ell_{AB}=\mathcal{O}(\tilde{a}/\tilde{\ell})$,
\begin{equation}
\int_{\mathcal{C}_{AB}}\mathbf{J}\cdot\mathrm{d}\boldsymbol{x}=\mathcal{O}(\tilde{a}/\tilde{\ell}).
\label{eq:order_int}
\end{equation}
\eqref{eq:order_int} shows that the LHS of \eqref{eq:int_J} has no $\mathcal{O}(1)$ terms. On the other hand, in the reduced-order model leading to \eqref{Seq:avg_pnp}, we solve for the dynamics of the leading-order charge density and potential in the limit of slender pores. Therefore, collecting terms on the expansions of both sides of \eqref{eq:int_J}, the $\mathcal{O}(1)$ terms of the RHS have to cancel out. For points $A$ and $B$ that pertain to the ends of the arbitrary pores $i$ and $j$, respectively, we find
\begin{equation}
\varphi_i|_{z_i(A)}=\varphi_j|_{z_j(B)}.
\label{eq:tr_bcs_pore}
\end{equation}
Since the lengths of the transition regions have been neglected in the current framework, the paths $\mathcal{C}_{AB}$ connecting pore centerlines are effectively mathematically treated as points. \eqref{eq:tr_bcs_pore} thus reveals the discontinuity of charge density and potential across connections of pores of different sizes -- physically, these properties can change quickly over the path of neglected length as long as the path is sufficiently long that profiles with radial equilibrium form in the connected ends. In fact, substituting \eqref{eq:rhophi_to_rhobar} into \eqref{eq:tr_bcs_pore}, we find that
\begin{equation} 
\mathcal{D}_i\bar{\rho}_i(z_i(A),t)=\mathcal{D}_j\bar{\rho}_j(z_j(B),t),
\end{equation}
such that the charge density is indeed discontinuous when $\mathcal{D}_i\ne\mathcal{D}_j$, i.e., when pore sizes are different, in this mathematical representation of the connections as points. Correspondingly, since \eqref{eq:tr_bcs_pore} establishes the conservation of the sum of charge density and electric potential across a connection, a discontinuity of charge density implies the discontinuity of electric potential as well. This is a hindrance for a transmission line representation of the charging dynamics, given that the discontinuity of potential would require additional circuit elements to represent the connections. As discussed in the main text, we bypass this difficulty by constructing the transmission line representation in terms of the electrochemical potential of charge, which is continuous across connections.

In addition to the flux-matching boundary condition, we impose current conservation at the ends of the asymptotically thin transition regions, i.e., $\sum\limits_{i\in \mathrm{ junction}}\int_{A_i}\mathbf{J}_i(z_i(\mathrm{junction})))\cdot \hat{\boldsymbol{n}}_i\,\mathrm{d}A_i=0$, where $A_i$ is the cross-sectional area of the $i$-th intersecting pore and $\hat{\boldsymbol{n}}_i$ its unit normal vector pointing outward from the junction volume. In terms of the electrochemical potential of charge, constant over each cross-section, we obtain
\begin{equation}
\sum\limits_{i\in \mathrm{ junction}}A_i\hat{\boldsymbol{n}}_i\cdot\nabla_i\varphi_i|_{z_i(\mathrm{junction})}=0
\end{equation}
for each pore junction.

\subsubsection{Static Diffusion Layers}

Consider an arbitrary pore indexed by $i$ that is connected to the reservoir, with a reference potential $\phi=0$, through an SDL with a dimensionless length $\ell_{s,i}$. Assuming a one-dimensional potential profile in the electroneutral SDL, the flux-matching condition yields 
\begin{equation}    
\varphi_i(z_i=0,t)=-E_{s,i}(t)\ell_{s,i},
\label{eq:varphi_sdl}
\end{equation}
where $E_{s,i}$ is the uniform electric field in the SDL. Current conservation yields 
\begin{equation}
-A_i\dfrac{\partial\varphi_i}{\partial z_i}\bigg|_{z_i=0}=A_{s,i}E_{s,i},
\label{eq:current_sdl}
\end{equation}
where the mouth of the $i$-th pore is parameterized by $z_i=0$ and where $A_{s,i}$ is the cross-sectional area of the $i$-th SDL. Substituting \eqref{eq:varphi_sdl} into \eqref{eq:current_sdl},
\begin{equation}
    \dfrac{\partial\varphi_i}{\partial z_i}\bigg|_{z_i=0}=\mathrm{Bi}_i\,\varphi_i(z_i=0,t),
\end{equation}
where $\mathrm{Bi}_i=A_{s,i}/(A_i\ell_{s,i})$ is a Biot number analogue, in this case a ratio of resistances to charge transfer in the $i$-th pore and its corresponding SDL.

\subsubsection{Blocking Electrodes}

We assume blocking surface conditions, such that at the end of each closed pore, parameterized by $z_i=\ell_i$, the average charge flux vanishes, i.e.,
\begin{equation}
\dfrac{\partial \varphi_i}{\partial z_i}\bigg|_{z_i=\ell_i}=0.
\end{equation}

\subsubsection{Initial Conditions}

Initially, the dissociated salt forms an electroneutral solute, such that
\begin{equation}
    \bar{\rho}_i(z_i,t=0)=0,\quad z_i\in [0,\ell_i].
\end{equation}
From \eqref{eq:rhophi_to_rhobar} or its slit-pore equivalent, we find that the initial modified potential is given by
\begin{equation}
\varphi_i(z_i,t=0)=1,\quad z_i\in [0,\ell_i].
\end{equation}

\section{An Analytical Solution for a Junction with an Arbitrary Number of Dead-End Pores}

\begin{figure*}[ht!]
    \centering
    \includegraphics[width=\textwidth]{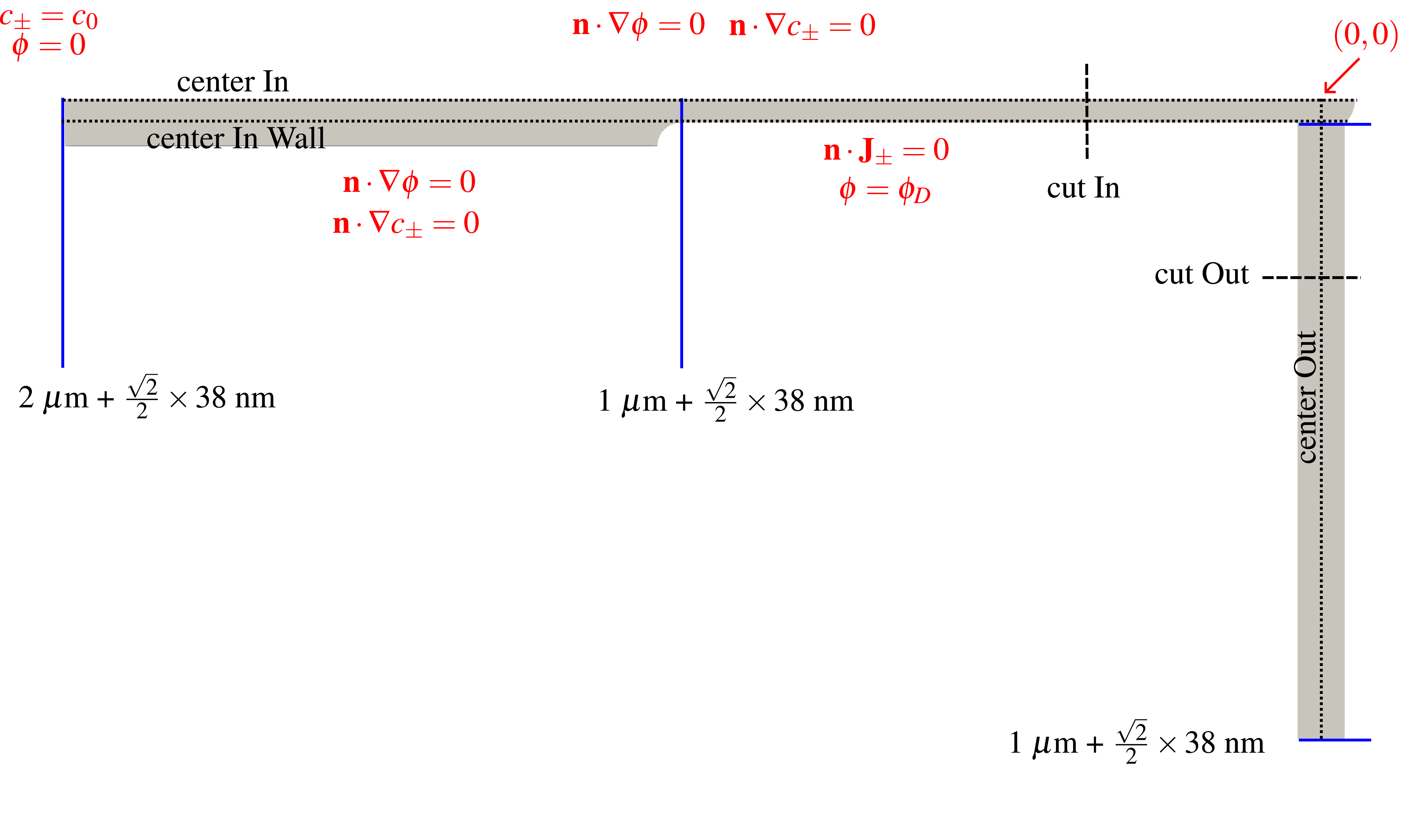}
    \caption{\textbf{Boundary conditions of the Direct Numerical Simulations}. The geometry of the electrolyte domain in the Y-junction is shown in gray. Due to top-down symmetry about the collinear axes of the static diffusion layer and the inlet pore, only half of the domain was simulated, using symmetry boundary conditions.}
    \label{fig:dns}
\end{figure*}

\begin{figure}
    \centering
    \includegraphics[width=\textwidth]{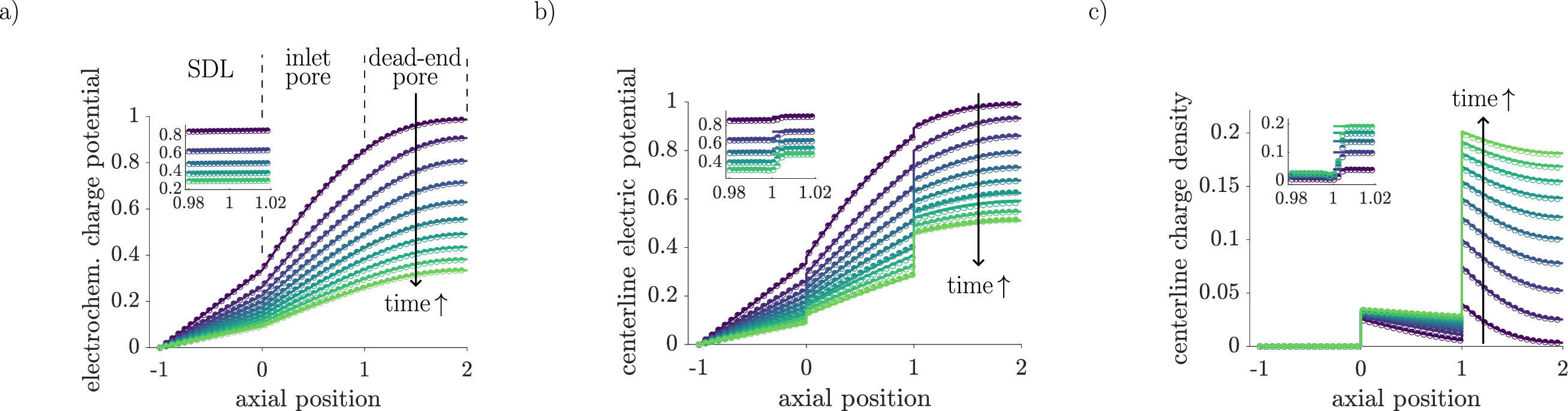}
    \caption{\textbf{Comparison of the analytical solution of the TL equations (solid lines) with DNS of the PNP equations (half-filled orbs) for a Y-junction of pores of equal lengths}. The axial (centerline) position is parameterized by the dimensionless arclength $s$. $-1<s<0$ in the SDL, $0<s<1$ in the inlet pore, and $1<s<2$ in each dead-end pore. The inset shows a zoom-in of the junction (near $s=1$). a) electrochemical potential of charge $\varphi$ as a function of the axial position. b) centerline electric potential $\phi_i(r_i=0,z_i,t)$ as a function of the axial position. c) centerline charge density $\rho_i(r_i=0,z_i,t)$ as a function of the axial position. While the electrochemical potential of charge changes smoothly across connections, the charge density and potential present sharp variations over lengths on the order of the Debye length. The magnitude of these changes across connections is well predicted by the TL model.}
    \label{fig:profiles}
\end{figure}

In this section, we discuss a method that can be used to solve our reduced-order equations for arbitrary pore networks. To demonstrate its use, we apply it to a single junction connecting one inlet pore to an arbitrary number $n$ of identical dead-end pores, shown in Fig. \ref{fig:profiles}a. We compare this solution for the case $n=2$ to full direct numerical simulations of the PNP equations to validate the transmission line model -- \eqref{Seq:avg_pnp} with the initial and boundary conditions derived in Sec. \ref{sec:ics_bcs}. In the setup considered, the first pore of relative pore size $\kappa_1$ and length $\ell_1$ connects to the reservoir through a static diffusion layer with a charge transfer Biot number $\mathrm{Bi}$, then branches out into $n$ identical dead-end pores with relative pore sizes $\kappa_2=\kappa_3=\cdots=\kappa_{n+1}$ and lengths $\ell_2=\ell_3=\cdots=\ell_{n+1}$. Using the symmetry of the dead-end pores, we only need to solve for one of them. Details of the geometry of the direct numerical simulations are shown in Fig. \ref{fig:dns}.

\subsection{General Solution}
\label{sec: soln}

We parameterize the mouth of the first pore, in contact with an SDL, by $z_1=0$. The axial coordinates of the pore extremities connected by the junction are $z_1=\ell_1$, $z_2=z_3=\cdots=z_{n+1}=0$. The blocking ends of the second and third pores are parameterized by $z_2=z_3=\cdots=z_{n+1}=\ell_2$. Due to symmetry, we only write the governing equations and initial or boundary conditions of the inlet pore and the first dead-end pore,
\begin{equation}
\dfrac{\partial\varphi_i}{\partial t}=\mathcal{D}_i\dfrac{\partial^2\varphi_i}{\partial z_i^2},\quad i=1,2,
\label{eq:gov_3pores}
\end{equation}
with initial conditions
\begin{equation}
    \varphi_i(z_i,t=0)=1,\quad i=1,2.
    \label{eq:ic}
\end{equation}
The SDL boundary condition is
\begin{equation}
\dfrac{\partial\varphi_1}{\partial z_1}\bigg|_{z_1=0}=\mathrm{Bi}\,\varphi_1(z_1=0,t),
\end{equation}
while the flux-matching boundary condition gives
\begin{equation}
\varphi_1(z_1=\ell_1,t)=\varphi_2(z_2=0,t).
\end{equation}
The current balance, using again the symmetry of the dead-end pores, yields
\begin{equation}
A_1\dfrac{\partial \varphi_1}{\partial z_1}\bigg|_{z_1=\ell_1}=nA_2\dfrac{\partial \varphi_2}{\partial z_2}\bigg|_{z_2=0},
\end{equation}
and the blocking-end boundary condition is
\begin{equation}
    \dfrac{\partial \varphi_2}{\partial z_2}\bigg|_{z_2=\ell_2}=0.
\end{equation}

We propose solutions by separation of variables in each pore of the form
\begin{equation}
\varphi_i(z_i,t)=\mathcal{Z}_i(z_i)T_i(t),\quad i=1,2,
\label{eq:sep_var_ansatz}
\end{equation}
allowing for different spatial dependencies in each region, but which are constrained by the boundary conditions. For distinct inlet and dead-end relative pore sizes, the difference in effective diffusivities $\mathcal{D}_i$ across distinct pores, along with the possibility of different total inlet and outlet areas in the connection, can break the orthogonality of the spatial basis functions, so we follow the procedure of Ref. \cite{tittle1965boundary} to build a set of orthogonal basis functions from $\mathcal{Z}_i$. This procedure has also been applied to the solution of diffusion equations in composite media \cite{pontrelli2007mass}. With \eqref{eq:sep_var_ansatz}, \eqref{eq:gov_3pores} implies
\begin{equation}
    \dfrac{\mathcal{Z}_i''}{\mathcal{Z}_i}=\dfrac{1}{\mathcal{D}_i}\dfrac{T_i'}{T_i}=-\lambda_i^2
    \label{eq:sep_var}
\end{equation}
where ordinary derivatives in either variable are represented by primes. $\lambda_i$ is the eigenvalue of the $i$-th pore, and the boundary conditions are
\begin{subequations}
\begin{equation}
\mathcal{Z}'_1(z_1=0)=\mathrm{Bi}\,\mathcal{Z}_1(z_1=0),
\end{equation}
\begin{equation}
\mathcal{Z}_1(z_1=\ell_1)=\mathcal{Z}_2(z_2=0),
\end{equation}
\begin{equation}
A_1\mathcal{Z}'_1(z_1=\ell_1)=nA_2\mathcal{Z}'_2(z_2=0),
\end{equation}
\begin{equation}
\mathcal{Z}'_2(z_2=\ell_2)=0.
\label{eq:blocking_end}
\end{equation}
\label{Seq:bcs}
\end{subequations} 

General solutions of the spatial dependencies in \eqref{eq:sep_var} are written as
\begin{subequations}
\begin{equation}
\mathcal{Z}_1(z_1)=\sin(\lambda_1 z_1)+\mathcal{B}_1\cos(\lambda_1 z_1),
    \label{eq:Z1}
\end{equation}
\begin{equation}    
\mathcal{Z}_2(z_2)=\mathcal{A}_2\sin(\lambda_2 (z_2-\ell_2))+\mathcal{B}_2\cos(\lambda_2 (z_2-\ell_2)),
    \label{eq:Z2}
\end{equation}
\end{subequations}
up to a constant multiplying \eqref{eq:Z1}, which will be included later in linear combinations of the ansatz of \eqref{eq:sep_var_ansatz} to satisfy the initial conditions. For the time dependency imposed by \eqref{eq:sep_var}, we obtain
\begin{equation}
    T_i(t)=\exp(-\lambda_i^2\mathcal{D}_it),\quad i=1,2,
\end{equation}
again up to a constant dependent on the initial conditions, which will be included later. In order to satisfy a common time dependency (see Refs. \cite{tittle1965boundary,pontrelli2007mass}) for all of the pores, we require
\begin{equation}
\lambda_1^2\mathcal{D}_1=\lambda_2^2\mathcal{D}_2.
\label{eq:lambda_1_2}
\end{equation}
Furthermore, applying the boundary conditions in \eqref{Seq:bcs}, we find
\begin{subequations}
\begin{equation}
    \mathcal{A}_2=0,
\end{equation}
\begin{equation}
\mathcal{B}_1=\mathrm{Bi}^{-1}\lambda_1,
\end{equation}
\begin{equation}
\mathcal{B}_2=\dfrac{\sin(\lambda_1\ell_1)+\mathrm{Bi}^{-1}\lambda_1\cos(\lambda_1\ell_1)}{\cos(\lambda_2\ell_2)},
\label{eq:bcs_c}
\end{equation}
\begin{equation}
\lambda_1A_1[\cos(\lambda_1\ell_1)-\mathrm{Bi}^{-1}\lambda_1\sin(\lambda_1\ell_1)]-n\mathcal{B}_2\lambda_2A_2\sin(\lambda_2\ell_2)=0.
\label{eq:bcs_d}
\end{equation}
\end{subequations}
The combination of \eqref{eq:lambda_1_2}, \eqref{eq:bcs_c}, and \eqref{eq:bcs_d} yields the characteristic equation
\begin{equation}
    \dfrac{A_1}{\sqrt{\mathcal{D}_1}}[1-\mathrm{Bi}^{-1}\lambda_1\tan(\lambda_1\ell_1)]-\dfrac{nA_2}{\sqrt{\mathcal{D}_2}}[\mathrm{Bi}^{-1}\lambda_1+\tan(\lambda_1\ell_1)]\tan(\lambda_1\ell_2\sqrt{\mathcal{D}_1/\mathcal{D}_2})=0.
    \label{eq:char_eqn}
\end{equation}
It should be noted that a family of eigenvalues $\lambda_{1k}$, $k\in \mathbb{N}$, satisfies this characteristic equation, which we enumerate by the additional index $k$.

As previously discussed, due to differences in the effective diffusivities $\mathcal{D}_i$ or in the total inlet and outlet areas of the connection, the basis functions $\mathcal{Z}_{ik}$ -- where the index $k$ identifies which eigenvalue $\lambda_{ik}$ corresponds to this $i$-th pore basis function -- may not be orthogonal under the standard integral inner product. We follow the method proposed by Ref. \cite{tittle1965boundary} to perform a Gram-Schmidt orthogonalization of the family of basis functions $\mathcal{Z}_{ik}(z_i)$. To this end, we multiply the basis function of each pore by a coefficient $\mathfrak{C}_i$ that ensures the orthogonality of functions corresponding to distinct eigenvalues, i.e., we seek $\mathfrak{C}_i$ such that
\begin{equation}
    (\lambda_{1k}^2-\lambda_{1l}^2)[\mathfrak{C}_1^2\int_0^{\ell_1}\mathcal{Z}_{1k}(z_1)\mathcal{Z}_{1l}(z_1)\,dz_1+\mathfrak{C}_2^2\int_0^{\ell_2}\mathcal{Z}_{2k}(z_2)\mathcal{Z}_{2l}(z_2)\,dz_2]=0,\quad k\ne l,\quad k,l\in\mathbb{N}.
    \label{eq:orthogonality}
\end{equation}
In this case, a single coefficient is required to enforce orthogonality, such that we are free to choose $\mathfrak{C}_1=1$. Using \eqref{eq:sep_var} and the boundary conditions in \eqref{Seq:bcs} to integrate by parts, we find that $\mathfrak{C}_2^2=nA_2\mathcal{D}_1/(A_1\mathcal{D}_2)$. Consequently, the squared norm of the set of orthogonalized basis functions, defined by
\begin{equation}
N_k^2=\int_0^{\ell_1}\mathcal{Z}_{1k}^2(z_1)\,dz_1+\mathfrak{C}_2^2\int_0^{\ell_2}\mathcal{Z}_{2k}^2(z_2)\,dz_2,
\label{eq:norm_sq}
\end{equation}
is given by
\begin{align}
    N_k^2&=\dfrac{1}{2}\left[(1+\mathrm{Bi}^{-2}\lambda_{1k}^2)\ell_1+\mathrm{Bi}^{-1}-\mathrm{Bi}^{-1}\cos(2\lambda_{1k}\ell_1)+\dfrac{\mathrm{Bi}^{-2}\lambda_{1k}^2-1}{2\lambda_{1k}}\sin(2\lambda_{1k}\ell_1)\right]+\nonumber\\ &\dfrac{nA_2\mathcal{D}_1}{2A_1\mathcal{D}_2}\dfrac{[\sin(\lambda_{1k}\ell_1)+\mathrm{Bi}^{-1}\lambda_{1k}\cos(\lambda_{1k}\ell_1)]^2}{{\cos^2(\lambda_{1k}\ell_2\sqrt{\mathcal{D}_1/\mathcal{D}_2})}}\left[\ell_2+\sqrt{\dfrac{\mathcal{D}_2}{\mathcal{D}_1}}\dfrac{\sin(2\lambda_{1k}\ell_2\sqrt{\mathcal{D}_1/\mathcal{D}_2})}{2\lambda_{1k}}\right].
\end{align}
We expand the initial condition in the region corresponding to each pore in terms of the quasi-orthogonal set of functions $\mathcal{Z}_{ik}(z_i)$, and the Fourier coefficients may be calculated and summed over all pore domains, yielding an expansion
\begin{equation}
\varphi_i(z_i,t)=\sum_{k=1}^\infty \mathcal{A}_k\mathcal{Z}_{ik}(z_i)\exp(-\lambda_{1k}^2\mathcal{D}_1t),
    \label{eq:fourier_soln}
\end{equation}
solely determined by the sequence of coefficients $\mathcal{A}_k$, which are found from the initial condition, \eqref{eq:ic}. In fact, multiplying the initial electrochemical potential of charge of the $i$-th pore by $\mathfrak{C}_i^2\mathcal{Z}_{ik}(z_i)$, integrating over the centerlines using the orthogonality relation in \eqref{eq:orthogonality} and the norm squared shorthand calculated in \eqref{eq:norm_sq}, we find
\begin{equation}
    \mathcal{A}_k=\dfrac{1}{N_k^2}\sum_{i=1}^N\mathfrak{C}_i^2\int_0^{\ell_i} \mathcal{Z}_{ik}(z_i)dz_i.
\end{equation}
Evaluating these integrals, we find
\begin{align}
    \label{eq:Ak}
    \mathcal{A}_k=\dfrac{1}{\lambda_{1k}N_k^2}&\left\{1-\cos(\lambda_{1k}\ell_1)+\mathrm{Bi}^{-1}\lambda_{1k}\sin(\lambda_{1k}\ell_1)+\right.\\ &\left.
    \dfrac{nA_2\sqrt{\mathcal{D}_1}}{A_1\sqrt{\mathcal{D}_2}}\left[\sin(\lambda_{1k}\ell_1)+\mathrm{Bi}^{-1}\lambda_{1k}\cos(\lambda_{1k}\ell_1)\right]\tan(\lambda_{1k}\ell_2\sqrt{\mathcal{D}_1/\mathcal{D}_2})\right\}.\nonumber
\end{align}

A comparison of the analytical solution of \eqref{eq:fourier_soln}--\eqref{eq:Ak} with a direct numerical simulation of the full PNP equations is shown in Fig. \ref{fig:dns}. Details of the profiles in the junctions are shown. The electrochemical potential of charge presents a smooth variation, while both the electric potential and the charge density change sharply across the connection.

\subsection{Physical Insights from the Analytical Solution}

The solution method used in Sec. \ref{sec: soln} is applicable to arbitrary networks that are representable by the transmission line model. In this section, we discuss the insights that the form of the solution gives to the double-layer charging physics of the model. We center the discussion around the significance of the eigenvalues $\lambda_{1k}$ of the general solutions, of the form of \eqref{eq:fourier_soln}, to physical properties of interest, namely the charging timescale and power density of the system. Then, we discuss the specific takeaways from the charging dynamics of a junction, calculated in the previous subsection, to the charging dynamics of porous networks. We also address possible shortcomings of extrapolating the takeaways of this toy-model geometry to general networks.

As a result of the one-dimensional heat equation form of the transmission line \eqref{Seq:avg_pnp}, solutions of the form of \eqref{eq:fourier_soln} hold in general. Therefore, with the linear homogeneous boundary conditions, long-time exponential decay of the electrochemical potential of charge is a universal feature of the model. The decay timescale is dictated by the reciprocal of the coefficient of the slowest decaying exponential, $\tau=1/(\lambda_{i1}^2\mathcal{D}_i)$, where $\lambda_{i1}$ is the lowest eigenvalue corresponding to the $i$-th pore. Note that it does not matter which pore is chosen for the analysis since all pores have a common time dependency; see \eqref{eq:lambda_1_2}.

The electrode power density can be calculated as the ratio of its energy density to its charging timescale. Since the applied potential dependency is factored out in the dimensionless formulation, the dimensionless electrode power density is given by the ratio of capacitance to charging timescale. The pore network geometry dictates the volumetric capacitance through the effective diffusivity-dependent steady-state charge densities of the pores. Furthermore, it determines the constants $\mathcal{A}_k$, the form functions $\mathcal{Z}_{ik}(z_i)$, and the eigenvalues $\lambda_{ik}$ through the boundary conditions and effective diffusivities, thus setting the long-time charging timescale of the medium. For a network of pores of equal lengths and sizes, the electrode volumetric capacitance is 
\begin{equation}
 \sum\limits_{i=1}^NC_i=\dfrac{p}{\mathcal{D}},   
\end{equation}
where $\mathcal{D}$ is the common effective diffusivity and $p$ is the porosity. Since this capacitance is itself inversely proportional to the effective diffusivity (see also Fig. 1c), this yields the electrode volumetric power density
\begin{equation}    
\mathcal{P}=\dfrac{\sum_{i=1}^NC_i}{\tau}= p\lambda_{11}^2.
\end{equation}
This result means that there are two geometrical contributions of distinct natures to the power density of a monodisperse pore network. The total pore volume affects the dimensionless capacitance of the network. On the other hand, for a monodisperse pore size distribution, the lowest eigenvalue $\lambda_{11}$
is exclusively dependent on the geometry of the network through connections, which introduce divisions of current or alter the lengths of the paths of the ions. $\lambda_{11}$ has no dependency on cross-sections or effective diffusivities. In fact, only the ratios of cross-sectional areas and effective diffusivities of distinct pores enter the characteristic equation, either through the boundary conditions or the common time dependencies of the form of \eqref{eq:lambda_1_2}. 
Finally, it should be noted that the power density is not dependent on the effective diffusivity of the pores, since it affects commensurately the capacitance and charging timescale.

We can better appreciate the effect of the geometry on the eigenvalues, and thus on the power density of the system, through the application to the junction problem solved in Sec. \ref{sec: soln}. Recall that the purpose of this junction model is twofold: first, it is used as a simple geometry for validation against direct numerical simulations of the PNP in the main text; and second, to serve as a minimal model of the effects of the coordination number in a connection, i.e., to capture the essential physics of the addition of pores to a connection. In this section, we focus on the latter. To simplify the problem further, let us assume that all pores are identical and that the SDL has a negligible resistance to charge transfer ($\mathrm{Bi}\to\infty$). In this case, the characteristic \eqref{eq:char_eqn} simplifies to
\begin{equation}
    1-n\tan^2(\lambda_{i})=0.
    \label{eq:char_eqn_simplified}
\end{equation}
The lowest eigenvalue is found from \eqref{eq:char_eqn_simplified}. Its reciprocal squared, proportional to the charging timescale, is plotted in Fig. 3e and compared to the single-pore result $\ell_i^2/\lambda_{i1}^2=16\ell_i^2/\pi^2$ (for $n=1$, the two pores can be conceived as a single longer pore with length $\ell_i$). Note that for $n\gg 1$, \eqref{eq:char_eqn_simplified} yields $\lambda_{i1}\sim 1/\sqrt{n}$, resulting in $\tau\sim n$. This behavior may be counterintuitive since the total length of the network increases quadratically with the number of dead-end pores, but the charging timescale increases only linearly with it. This result indicates that the primary effect of the additional dead-end pores is not to increase the length traversed by the charge, but to provide simultaneous alternative paths of the same length and to divide the current going through each one. Similar behavior can be expected in a regular monodisperse lattice, e.g. in the plots of Figs. 4 and \ref{fig:mono}, where the approximate periodicity (neglecting lattice edge effects) in the direction normal to charging precludes the ions from being transported into different rows and thus traversing lengths of higher order than the side of the lattice. We speculate that more disordered porous media could present stronger dependencies of the eigenvalues on the number of connections due to the possibility of transport being dominated by tortuous paths.

\section{Effects of Connectivity}

\begin{figure*}[ht!]
    \centering
    \includegraphics[width=\textwidth]{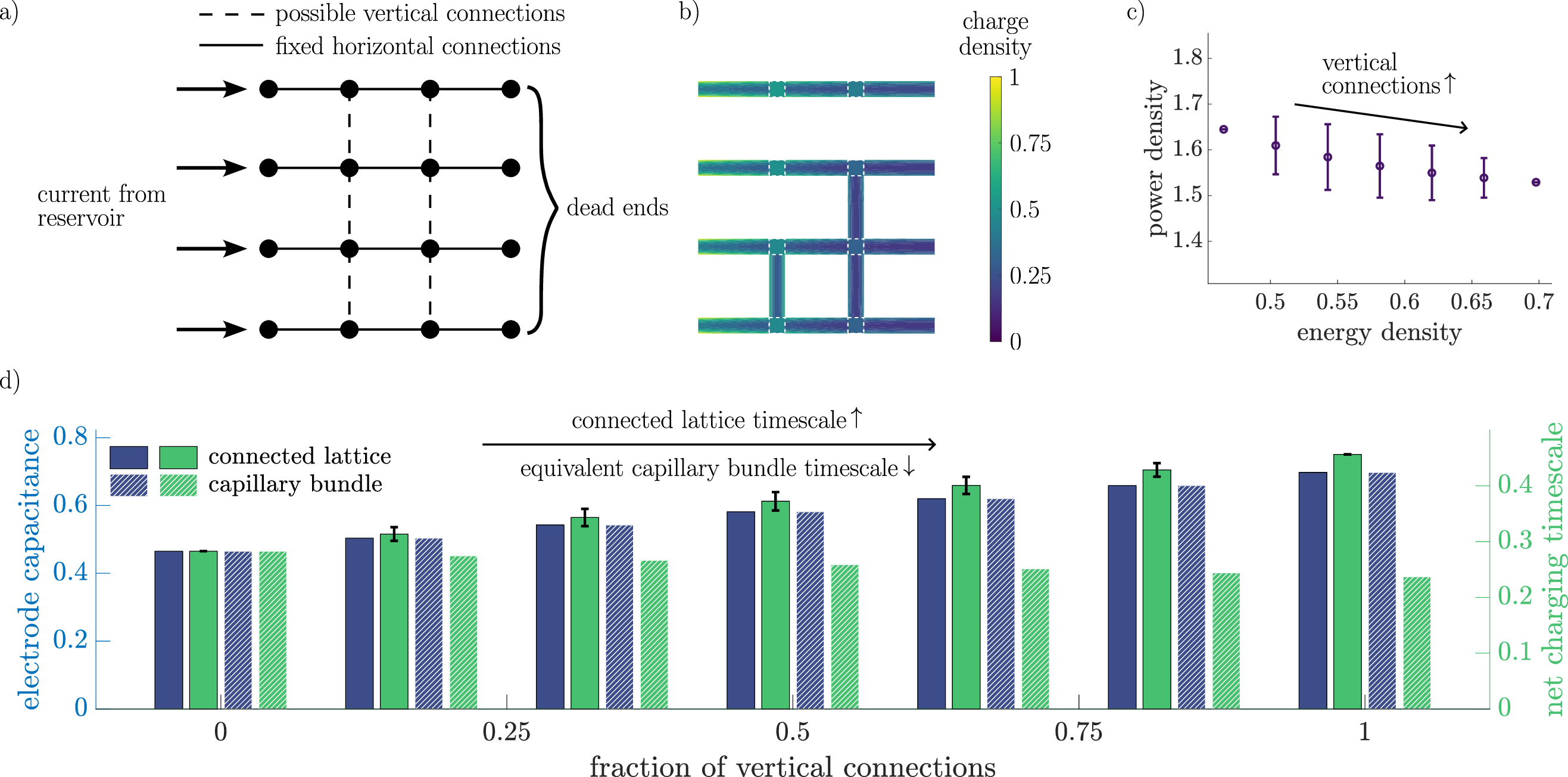}
    \caption{\textbf{Effects of Connectivity in the Charging Characteristics of Lattices.} a) Schematic of the $4\times 4$ lattice geometry studied. Sites, the filled orbs, are pore inlets, connections, or dead ends. Solid lines are fixed horizontal pores, such that the lattice is above the percolation threshold. We study the effect of the presence of vertical pores that connect the horizontal pores. Dashed lines indicate all the possible vertical pores. Current flows from the reservoir to the leftmost sites (for $\phi_D<0$). The rightmost sites are blocking ends, such that no current flows through them. b) Contour plot of charge density of one possible configuration with a fraction of vertical connections $f_y=1/2$ at time $t=2.5$. The thin transition regions enclosed by dashed white square borders are not resolved by the model, the artificially introduced color corresponds to the average charge density of the sides of each transition region. c) Ragone plot -- dimensionless electrode power density vs. energy density -- for all fractions of vertical connections $f_y\in \{0,1/6,1/3,1/2,2/3,5/6,1\}$. Error bars in parts c) and d) are given by the standard deviation of distinct configurations with the same number of vertical pores. d) Bar plot of electrode volumetric capacitance (left y-axis) and net charging timescale (right y-axis) vs. the fraction of vertical connections in the lattice for the connected lattice (plain color) and an equivalent capillary bundle (white hatched fill), i.e., one with no vertical pores, but with the same number of horizontal pores and the same capacitance. The plot shows that despite the increase in energy density due to the inclusion of additional pores within the same electrode volume, the charging timescale increase is even higher, promoting a slight decrease in the power density. However, since the equivalent capillary bundle of a higher $f_y$ must have pores with larger relative sizes, its charging timescale decreases with $f_y$.}
    \label{fig:mono}
\end{figure*}

To illustrate and quantify the effects of connectivity in the charging porous networks, we investigate charging dynamics of the $4\times 4$ lattices of identical pores illustrated in Fig. \ref{fig:mono}a. Here, the pore surfaces of the lattice are held at constant potential $\phi_D$, and hence the current flows from left to right (for $\phi_D<0$), i.e., from the reservoir into the porous network. The sites denoted by circles are either pore inlets, dead ends, or junctions. The horizontal pores, indicated by solid lines, are always present such that the lattice is above the percolation threshold. Different configurations are achieved through possible insertions of vertical pores at the locations represented by dashed lines. The connectivity is controlled through the number of vertical pores in the lattice, which we quantify by the fraction of vertical connections $f_y$, i.e., the number of vertical pores in a given configuration over the maximum possible number of vertical pores. As shown in the main text (see Fig. 3), in addition to the number of vertical connections, their placement also plays a significant role in the network charging dynamics. Therefore, we determine the charging characteristics of a given number of vertical connections through ensemble averages over all such possible configurations.

As per the methodology described above, even the $4 \times 4$ lattice has $64$ distinct configurations.  A contour plot of charge density for a configuration with $f_y=1/2$ is provided in Fig. \ref{fig:mono}b. The junctions, bounded by white dashed squares, are not resolved by the model and are thus colored according to the average charge densities of the sides for visual clarity; see details in Materials and Methods. The qualitative effect of the connections becomes apparent from the contours of the last pores of each row. The dead-end pore of the first (topmost) row, i.e., the one with no vertical connections, has a larger charge density than the dead-end pores of the remaining rows. This occurs because additional connections create alternative pathways for the ions, dividing the current that goes through each pore and delaying the charging process.

While the addition of pores delays the charging process, it also adds pore volume, thus increasing the capacitance of the system. Therefore, it is necessary to quantify the competition between the increase in capacitance and the delay in the charging timescale. In this setting, a property of relevance for technological applications is the electrode power density. Fig. \ref{fig:mono}c shows that the increase in volumetric capacitance, and therefore energy density, with the fraction of vertical connections is accompanied by a mild decrease in the power density. This trend is consistent with that of the junction described in Sec. \ref{sec: soln}, as shown in Fig. 3f. We note that the error bars present demonstrate the difference that can arise due to the placement of the pores, even if the value of $f_y$ is identical. 

Another way to quantify the impact of pore connections is to compare a given connection with an equivalent capillary bundle that has identical capacitance per unit volume.  This comparison allows one to focus solely on the impact of connections on the charging timescale since the two geometries have equal electrode capacitance. We illustrate this by comparing electrode (volumetric) capacitance (left y-axis) and net charging timescale (right y-axis) vs. the fraction of vertical connections in Fig. \ref{fig:mono}d for the two geometries, i.e., connected lattice and an equivalent capillary bundle. For a connected lattice, both the capacitance and the charging timescale increase with the number of connections. However, the charging timescale increases more than the volumetric capacitance when connections are added, which is consistent with trends reported in Fig. \ref{fig:mono}c. In contrast, the equivalent capillary bundle displays a reduced timescale. This difference arises since capillaries with large pore sizes are required to match the electrode capacitance.  However, larger capillaries charge faster since the dominant mechanism is electromigration. Mathematically, the effective diffusion coefficient is boosted in the electromigration-dominated regime $\mathcal{D}_i \sim \kappa_i/2$, which subsequently reduces the timescale.

We would like to clarify that the results described in Fig. \ref{fig:mono}c are not to be extrapolated to porous networks with high porosity. The networks described above do not take into account volume restrictions on pore placement. For highly porous networks, the large pores might not be physically realizable due to these volume constraints. Therefore, we anticipate that there would be an optimal combination of average size and connectivity for a desired property set. Still, the results described above highlight the profound impact that pore connectivity has on the charging dynamics of an electrode. The presence of connections significantly slows down the charging process. We would like to note that this slowdown is observable even for a $4\times 4$ lattice and we anticipate that it will be enhanced for a larger density and porosity of pores, similarly to Fig. 3e and Ref. \cite{lian2020blessing}. As such, we envision that the proposed approach could shed light on the effects of tortuosity. For instance, the recent results outlined by Nguyen et al. \cite{nguyen2020electrode} suggest an electrode tortuosity factor. However, the analysis by the authors focuses only on the limit of thin double layers since they assumed surface capacitance only. By including our methodology, such an analysis can be extended.


\begin{thebibliography}{9}
\small
\bibitem{bazant2004diffuse}
M. Z. Bazant, K. Thornton, A. Ajdari, Diffuse-charge dynamics in electrochemical systems. Phys. Rev. E 70, 021506 (2004).

\bibitem{ho2005electrolytic}
C. Ho, et al., Electrolytic transport through a synthetic nanometer-diameter pore. Proc. Natl.
Acad. Sci. 102, 10445–10450 (2005).

\bibitem{muthukumar2006simulation}
M. Muthukumar, C. Kong, Simulation of polymer translocation through protein channels. Proc.
Natl. Acad. Sci. 103, 5273–5278 (2006).

\bibitem{davidson2014chaotic}
S. M. Davidson, M. B. Andersen, A. Mani, Chaotic induced-charge electro-osmosis. Phys. Rev.
Lett. 112, 128302 (2014).

\bibitem{jubin2018dramatic}
L. Jubin, A. Poggioli, A. Siria, L. Bocquet, Dramatic pressure-sensitive ion conduction in conical
nanopores. Proc. Natl. Acad. Sci. 115, 4063–4068 (2018).

\bibitem{simon2008materials}
P. Simon, Y. Gogotsi, Materials for electrochemical capacitors.
Nat. Mater. pp. 845–854
(2008).

\bibitem{simon2020perspectives}
P. Simon, Y. Gogotsi, Perspectives for electrochemical capacitors and related devices. Nat.
Mater. 19, 1151–1163 (2020).

\bibitem{sakaguchi2007charging}
H. Sakaguchi, R. Baba, Charging dynamics of the electric double layer in porous media. Phys.
Rev. E 76, 011501 (2007).

\bibitem{biesheuvel2010nonlinear}
P. Biesheuvel, M. Bazant, Nonlinear dynamics of capacitive charging and desalination by
porous electrodes. Phys. Rev. E 81, 031502 (2010).

\bibitem{mirzadeh2014enhanced}
M. Mirzadeh, F. Gibou, T. M. Squires, Enhanced charging kinetics of porous electrodes: Surface
conduction as a short-circuit mechanism. Phys. Rev. Lett. 113, 097701 (2014).

\bibitem{schmuck2015homogenization}
M. Schmuck, M. Z. Bazant, Homogenization of the poisson–nernst–planck equations for ion
transport in charged porous media. SIAM J. Appl. Math. 75, 1369–1401 (2015).

\bibitem{jiang2021large}
Y. Jiang, et al., Large-surface-area activated carbon with high density by electrostatic densifi-
cation for supercapacitor electrodes. Carbon 175, 281–288 (2021).

\bibitem{chu20213d}
T. Chu, S. Park, K. Fu, 3d printing-enabled advanced electrode architecture design. Carbon
Energy 3, 424–439 (2021).

\bibitem{kondrat2023theory}
S. Kondrat, G. Feng, F. Bresme, M. Urbakh, A. A. Kornyshev, Theory and simulations of ionic
liquids in nanoconfinement. Chem. Rev. (Washington, DC, U. S.) (2023).

\bibitem{wu2022understanding}
J. Wu, Understanding the electric double-layer structure, capacitance, and charging dynamics.
Chem. Rev. (Washington, DC, U. S.) 122, 10821–10859 (2022).

\bibitem{lahrar2021carbon}
E. H. Lahrar, P. Simon, and C. Merlet, Carbon–carbon supercapacitors: Beyond the average pore size or how electrolyte confinement and inaccessible pores affect the capacitance. J. Chem. Phys. 155.18 (2021).

\bibitem{kilic2007steric1}
M. S. Kilic, M. Z. Bazant, A. Ajdari, Steric effects in the dynamics of electrolytes at large applied
voltages. i. double-layer charging. Phys. Rev. E 75, 021502 (2007).

\bibitem{kilic2007steric2}
M. S. Kilic, M. Z. Bazant, A. Ajdari, Steric effects in the dynamics of electrolytes at large applied
voltages. ii. modified poisson-nernst-planck equations. Phys. Rev. E 75, 021503 (2007).

\bibitem{storey2012effects}
B. D. Storey, M. Z. Bazant, Effects of electrostatic correlations on electrokinetic phenomena.
Phys. Rev. E 86, 056303 (2012).

\bibitem{xu2014self}
Z. Xu, M. Ma, P. Liu, Self-energy-modified poisson-nernst-planck equations: Wkb approxima-
tion and finite-difference approaches. Phys. Rev. E 90, 013307 (2014).

\bibitem{gupta2020ionic}
A. Gupta, A. G. Rajan, E. A. Carter, H. A. Stone, Ionic layering and overcharging in electrical double
layers in a poisson-boltzmann model. Phys. Rev. Lett. 125, 188004 (2020).

\bibitem{de2020interfacial}
J. P. de Souza, Z. A. Goodwin, M. McEldrew, A. A. Kornyshev, M. Z. Bazant, Interfacial layering in
the electric double layer of ionic liquids. Phys. Rev. Lett. 125, 116001 (2020).

\bibitem{kondrat2014accelerating}
S. Kondrat, P. Wu, R. Qiao, A. A. Kornyshev, Accelerating charging dynamics in subnanometre
pores. Nat. Mater. 13, 387–393 (2014).

\bibitem{tomlin2022impedance}
R. J. Tomlin, T. Roy, T. L. Kirk, M. Marinescu, D. Gillespie, Impedance response of ionic liquids in
long slit pores. J. Electrochem. Soc. 169, 120513 (2022).

\bibitem{de1963porous}
R. De Levie, On porous electrodes in electrolyte solutions: I. capacitance effects. Electrochim.
Acta 8, 751–780 (1963).

\bibitem{de1964porous}
R. De Levie, On porous electrodes in electrolyte solutions—iv. Electrochim. Acta 9, 1231–
1245 (1964).

\bibitem{black2010pore}
J. M. Black, H. A. Andreas, Pore shape affects spontaneous charge redistribution in small pores.
J. Phys. Chem. C 114, 12030–12038 (2010).

\bibitem{movskon2021transmission}
J. Moškon, M. Gaberšcek, Transmission line models for evaluation of impedance response of
insertion battery electrodes and cells. J. Power Sources Adv. 7, 100047 (2021).

\bibitem{gebbie2013ionic}
M. A. Gebbie, et al., Ionic liquids behave as dilute electrolyte solutions. Proc. Natl. Acad. Sci.
110, 9674–9679 (2013).

\bibitem{bi2020molecular}
S. Bi, et al., Molecular understanding of charge storage and charging dynamics in supercapacitors with mof electrodes and ionic liquid electrolytes. Nat. Mater. 19, 552–558 (2020).

\bibitem{zeng2021modeling}
L. Zeng, et al., Modeling galvanostatic charge–discharge of nanoporous supercapacitors. Nat.
Comput. Sci. 1, 725–731 (2021).

\bibitem{janssen2021transmission}
M. Janssen, Transmission line circuit and equation for an electrolyte-filled pore of finite length.
Phys. Rev. Lett. 126, 136002 (2021).

\bibitem{gupta2020charging}
A. Gupta, P. J. Zuk, H. A. Stone, Charging dynamics of overlapping double layers in a cylindrical
nanopore. Phys. Rev. Lett. 125, 076001 (2020).

\bibitem{henrique2022charging}
F. Henrique, P. J. Zuk, A. Gupta, Charging dynamics of electrical double layers inside a cylindrical pore: predicting the effects of arbitrary pore size. Soft Matter 18, 198–213 (2022).

\bibitem{henrique2022impact}
F. Henrique, P. J. Zuk, A. Gupta, Impact of asymmetries in valences and diffusivities on the
transport of a binary electrolyte in a charged cylindrical pore. Electrochim. Acta 433, 141220
(2022).

\bibitem{lian2020blessing}
C. Lian, M. Janssen, H. Liu, R. van Roij, Blessing and curse: how a supercapacitor’s large
capacitance causes its slow charging. Phys. Rev. Lett. 124, 076001 (2020).

\bibitem{biesheuvel2011diffuse}
P. M. Biesheuvel, Y. Fu, M. Z. Bazant, Diffuse charge and faradaic reactions in porous electrodes.
Phys. Rev. E 83, 061507 (2011).

\bibitem{alexander2013fundamentals}
C. K. Alexander, M. N. O. Sadiku, Fundamentals of Electric Circuits. (McGraw-Hill), (2013).

\bibitem{deen1998analysis}
W. M . Deen, Analysis of Transport Phenomena. (Oxford University Press New York), (1998).

\bibitem{peters2016analysis}
P. B. Peters, R. Van Roij, M. Z. Bazant, P. M. Biesheuvel (2016). Analysis of electrolyte transport through charged nanopores. Phys. Rev. E, 93(5), 053108.

\bibitem{alizadeh2017multiscale}
S. Alizadeh , A. Mani (2017). Multiscale model for electrokinetic transport in networks of pores, part I: model derivation. Langmuir, 33(25), 6205-6219.

\bibitem{forse2017direct}
A. C. Forse, et al., Direct observation of ion dynamics in supercapacitor electrodes using in situ
diffusion nmr spectroscopy. Nat. Energy 2, 1–7 (2017).

\bibitem{newman1962theoretical}
J. S. Newman, C. W. Tobias, Theoretical analysis of current distribution in porous electrodes. J.
Electrochem. Soc. 109, 1183 (1962).

\bibitem{taocharging}
H. Tao, Z. Xu, C. Lian, R. van Roij, \& H. Liu, Charging dynamics in a laminate‐electrode model for graphene‐based supercapacitors. AIChE J., Advance Access.

\bibitem{presser2012electrochemical}
V. Presser, et al., The electrochemical flow capacitor: A new concept for rapid energy storage
and recovery. Adv. Energy Mater. 2, 895–902 (2012).

\bibitem{zhang2014highly}
C. Zhang, et al., Highly porous carbon spheres for electrochemical capacitors and capacitive
flowable suspension electrodes. Carbon 77, 155–164 (2014).

\bibitem{mo2023horn}
T. Mo, et al., Horn-like pore entrance boosts charging dynamics and charge storage of
nanoporous supercapacitors. ACS nano (2023).

\bibitem{pedersen2023equivalent}
C. Pedersen, T. Aslyamov, M. Janssen, Equivalent circuit and continuum modeling of the
impedance of electrolyte-filled pores. arXiv preprint arXiv:2305.02766 (2023).

\bibitem{gupta2018electrical}
A. Gupta, H. A. Stone, Electrical double layers: effects of asymmetry in electrolyte valence on
steric effects, dielectric decrement, and ion–ion correlations. Langmuir 34, 11971–11985
(2018).

\bibitem{jarvey2022ion}
N. Jarvey, F. Henrique, A. Gupta, Ion transport in an electrochemical cell: A theoretical framework to couple dynamics of double layers and redox reactions for multicomponent electrolyte
solutions. J. Electrochem. Soc. 169, 093506 (2022).

\bibitem{jarvey2023asymmetric}
N. Jarvey, F. Henrique, A. Gupta, Asymmetric rectified electric and concentration fields in multicomponent electrolytes with surface reactions. Soft Matter 19, 6032–6045 (2023).

\bibitem{fu2016graphene}
K. Fu, et al., Graphene oxide-based electrode inks for 3d-printed lithium-ion batteries. Adv.
Mat. 28, 2587–2594 (2016).

\bibitem{fleischmann2022continuous}
S. Fleischmann, et al., Continuous transition from double-layer to faradaic charge storage in
confined electrolytes. Nat. Energy 7, 222–228 (2022).

\bibitem{eymard2000finite}
R. Eymard, T. Gallouët, R. Herbin, Finite volume methods. Handb. Numer. Anal. 7, 713–1018
(2000).

\bibitem{versteeg2007introduction}
H. K. Versteeg, W. Malalasekera, An introduction to computational fluid dynamics: the finite
volume method. (Pearson education), (2007).

\bibitem{weller1998tensorial}
H. G. Weller, G. Tabor, H. Jasak, C. Fureby, A tensorial approach to computational continuum
mechanics using object-oriented techniques. Comput. Phys. 12, 620–631 (1998).

\bibitem{jasak2007openfoam}
H. Jasak, A. Jemcov, Z. Tukovic, et al., Openfoam: A c++ library for complex physics simulations in International Workshop on Coupled Methods in Numerical Dynamics. (IUC Dubrovnik
Croatia), Vol. 1000, pp. 1–20 (2007).

\bibitem{aslyamov2022relation}
T. Aslyamov, K. Sinkov, I. Akhatov, Relation between charging times and storage properties of nanoporous supercapacitors.
Nanomaterials 12, 587 (2022).

\bibitem{aslyamov2022analytical}
T. Aslyamov, M. Janssen, Analytical solution to the poisson–nernst–planck equations for the charging of a long electrolyte-filled slit pore. Electrochim. Acta 424, 140555 (2022).

\bibitem{huang2008theoretical}
J. Huang, B. G. Sumpter, V. Meunier, Theoretical model for nanoporous carbon supercapacitors. Angew. Chem. 120, 530–534
(2008).

\bibitem{chu2007surface}
K. T. Chu, M. Z. Bazant, Surface conservation laws at microscopically diffuse interfaces. J. Colloid Interface Sci. 315, 319–329
(2007).

\bibitem{tittle1965boundary}
C. Tittle, Boundary value problems in composite media: quasi-orthogonal functions. J. Appl. Phys. 36, 1486–1488 (1965).

\bibitem{pontrelli2007mass}
G. Pontrelli, F. de Monte, Mass diffusion through two-layer porous media: an application to the drug-eluting stent. Int. J.
Heat Mass Transf. 50, 3658–3669 (2007).

\bibitem{nguyen2020electrode}
T. T. Nguyen, et al., The electrode tortuosity factor: why the conventional tortuosity factor is not well suited for quantifying
transport in porous li-ion battery electrodes and what to use instead. Npj Comput. Mater. 6, 123 (2020).

\end{thebibliography}
\end{document}